\documentclass[final,3p,times,nopreprintline]{elsarticle}
\usepackage[numbers]{natbib}
\usepackage{url}
\providecommand{\keywords}[1]
{
  \small	
  \textbf{Keywords---} #1
}
\newcommand{\toright}[1]{\hspace*{\fill}{\footnotesize{#1}}}

\begin{document}

\markboth{Thomas Gehrmann, Bogdan Malaescu}{Precision QCD Physics at the LHC}

\title{\toright{{\footnotesize{ZU-TH 53/21}}}\\[0.5cm]
Precision QCD Physics at the LHC}

\author{{\large Thomas Gehrmann$^1$ and\  Bogdan Malaescu}$^2$ \\[2mm]
{$^1$Physik-Institut, Universit\"at Z\"urich, CH-8057 Z\"urich, Switzerland; email: thomas.gehrmann@uzh.ch} \\
{$^2$LPNHE, Sorbonne Universit\'e, Universit\'e Paris Cit\'e, CNRS/IN2P3, Paris, France, 75252; email: malaescu@in2p3.fr}}

\begin{abstract}
 This review describes the current status of precision QCD studies at the LHC. We introduce the main experimental and theoretical methods, discussing also their cross-stimulated developments and recent advances. The different types of QCD observables that are measured at the LHC, including cross-sections, event- and jet-level properties, for various final states, are summarised. Their relation to fundamental QCD dynamics and their impact on Standard Model parameter determinations are discussed on specific examples. The impact of QCD-related observables on direct and indirect searches for rare processes within and new physics beyond the Standard Model is outlined. 
\end{abstract}

\maketitle


\tableofcontents

\section{Introduction}

Quantum chromodynamics (QCD) is firmly established as the theory of the strong interactions~\cite{Gross:1973id,Politzer:1973fx}. Together with the theory of the electroweak interaction, 
it forms the Standard Model (SM) of particle physics, which has been validated through a multitude of experimental observations at collider 
and non-collider experiments
in the past decades.  Many aspects of the electroweak interaction have been experimentally tested to sub-per-cent level; 
when combined with theory predictions 
at a comparable level of precision, these measurements enable highly accurate extractions of electroweak parameters. With the increasing 
LHC data set and advances in QCD theory, measurements of QCD observables are now starting to attain a similar level of precision, thereby enabling highly 
accurate measurements of the QCD parameters and searching for small deviations from SM expectations as indirect probes of new physics. 
This review describes the current state of the art of precision QCD at the LHC,  by overviewing the body of existing measurements 
and associated theory predictions, and by discussing future directions and physics implications. 

Particles interacting through the strong interaction (partons: quarks and
gluons) are never observed as free particles, but only within bound 
states (hadrons: e.g.\ proton, neutron, pion). Free partons that are 
produced in the course of a high energy interaction fragment into 
a large number of hadrons, which are observed in the form of a jet 
emerging from the collision. In order to measure jet cross sections and 
to compare them with theoretical predictions, one has to specify a 
procedure to combine hadrons into jets (jet algorithm). Since the
jet algorithm analyses all final-state momenta in an event and compares 
them with a jet size parameter (often called jet radius), jet observables 
are inherently exclusive. 

Protons are not elementary particles, but composed of quarks and gluons;
consequently
QCD affects all observables at proton-proton colliders~\cite{esw,dissertori,campbell}. Physics studies 
at the LHC are therefore always relying on a good understanding of QCD 
effects, and many physics observables studied at the LHC also involve jets in 
the final state. The concept of QCD factorization allows to separate the 
dynamics at different resolution scales. The 
non-perturbative bound state dynamics of partons in 
the proton are described by parton distribution functions (PDFs), which
encode the momentum distributions of partons inside the proton and 
are determined through fits of theoretical predictions to experimental measurements. Any 
 hard scattering process is computed at the level of incoming 
 and outgoing partons, which subsequently fragment into 
hadrons. This hadron fragmentation is again governed by non-perturbative 
dynamics, and described by empirical models.

Quantitative predictions for hard processes in QCD are obtained by using perturbation 
theory, which expresses physical quantities as a power series in the respective coupling 
constants, and an increasing accuracy is obtained by  including higher orders in the perturbative 
expansion. The truncation of the perturbation series at a given order results in an uncertainty on the theory prediction, which is commonly quantified by varying the scales that are used for the renormalisation of the coupling constant and for the mass factorisation of the PDFs around some value that is characteristic for the process under consideration. 

The leading order (LO) normally captures only the gross features of an observable, 
inclusion of next-to-leading order (NLO) corrections is required to estimate the normalisation of 
the predictions, and even higher orders (NNLO, N3LO, $\ldots$) become mandatory if 
detailed event properties or per-cent level precision are in demand. 
In specific kinematical settings, especially if multiple 
kinematical scales are present in a single observable, these are supplemented by 
resummation of large logarithmic corrections to all orders in the coupling. 

The complexity of calculating an observable increases with each perturbative order. It proceeds through 
the evaluation of all scattering matrix elements, which derive from
particle scattering amplitudes (Feynman amplitudes) 
that contribute to a given final state at this order. The Feynman amplitudes 
can include both the exchange of intermediate virtual particles and the radiation of 
real final-state particles. To assemble the prediction for a specific final state, the momenta of all 
unobserved real and virtual particles have to be integrated over, often resulting in 
divergences arising from the regions of large (ultraviolet, UV) or small (infrared, IR) momenta in individual 
contributions. 
While the UV singularities are accounted for by renormalization in each contribution, the IR singularities cancel only in the sum 
of all contributions for an appropriately defined final state. Consequently, one requires techniques 
to extract IR singular terms and recombine them among the different contributions to an observable under consideration. The larger framework for these calculations is dimensional regularisation, which enables to make the singular contributions in loop and phase space integrals explicit in an analytic manner. The residues of these singularites are then computed either analytically or numerically, and the finite remainder is evaluated fully numerically. 

Cross sections and kinematical distributions at high energy colliders can depend on different scales, such as 
particle masses or transverse momenta. Consequently, the perturbative fixed-order coefficient functions 
can depend on the logarithms of scale-ratios. If the hierarchy between scales in an observable becomes large, these 
logarithms can deteriorate or even destroy the convergence of the perturbative series expansion. In order to 
obtain reliable predictions across the full kinematical range, these logarithmic corrections need to be resummed 
to all orders, and combined with the fixed-order corrections in a form that avoids double counting. 

Precision observables comprise the production of jets, photons, massive vector bosons $V=(W,Z)$ (observed through their 
lepton decay products) or heavy quarks.
Those final states are analysed as distributions in their kinematical 
variables. The experimental measurements are performed over a finite kinematical range, where 
the final state particles under consideration can be reconstructed reliably.  
The resulting fiducial cross sections are the objects that are ideally compared between theory and experiment, thereby requiring the 
theory predictions to take proper account of the specific kinematical restrictions of the corresponding measurements. 

Over the past fifteen years, the calculation of the scattering amplitudes relevant 
to NLO corrections both in QCD and 
the electroweak theory has been automated \cite{ellisrev}, and can be 
performed in a reliable and efficient manner up to high multiplicities. Calculations to this order 
are now done routinely in the framework of multi-purpose event simulation programs, such 
as SHERPA~\cite{sherpa}, HERWIG~\cite{herwig}, POWHEG~\cite{Alioli:2010xd} or mg5\_aMCatNLO~\cite{mg5}, which read in the 
tree-level and one-loop scattering amplitudes from automated tools~\cite{blackhat}
through standardised interfaces~\cite{blha}.  
These generic tools are complemented by 
the dedicated libraries of NLO processes MCFM~\cite{mcfm} and VBFNLO~\cite{vbfnlo}.

Owing to methodological advances in the calculation of two-loop amplitudes~\cite{laporta,gr,henn} and in the treatment of 
infrared singular real radiation~\cite{secdec,qtsub,ourant,stripper,njettiness,trocsanyi}, an increasing number of collider processes have been computed recently in fully differential form to
NNLO QCD accuracy~\cite{gudrunrev}.  
Following early results on Higgs 
production~\cite{qtsub,Anastasiou:2005qi} and the Drell-Yan process~\cite{fewz}, calculations at this order have been accomplished for all $2\to 2$ processes, like for 
example di-jet production~\cite{our2j,czakon2j}, vector-boson-plus-jet production~\cite{wjet,ourvj},
photon-plus-jet-production~\cite{gamj}
or top quark pair production~\cite{czakon1,grazzinitop}
and first NNLO calculations for 
$2\to 3$ processes were completed most recently~\cite{czakon3j}. 
All these calculations were implemented in the form of parton-level event generators, 
which provide 
full kinematical information on all final state particles, and consequently allow to account for the precise 
definition (jet algorithm, kinematical acceptance cuts) of observables used in the experimental analyses. Of these codes, only MATRIX (which computes for example vector boson 
pair production and top quark pair production at NNLO~\cite{grazzinitop,matrix}) 
and MCFM (for color-singlet final states,~\cite{mcfm2}) are 
available as open-source.

First steps towards N3LO calculations were taken most recently, with the 
computations of fully inclusive coefficient functions for Higgs production~\cite{hn3lo} and 
the Drell-Yan process~\cite{dyn3lo}, which are now being extended towards fully differential final states~\cite{hn3lodiff}. 

The resummation of leading logarithmic corrections (and of large parts of the first subleading logarithms) is 
accomplished by the parton shower approximation, which forms the basis of all modern multi-purpose 
Monte Carlo (MC) event 
generator programs:  PYTHIA~\cite{pythia},  HERWIG~\cite{herwig} and SHERPA~\cite{sherpa}. 
Resummation at higher logarithmic accuracy is accomplished by 
dedicated analytical methods, which require the calculation of various ingredients that take proper account 
of the logarithmic contributions from soft and collinear radiation processes and from initial-state radiation. Particularly relevant for
hadron collider observables is the resummation of large logarithms of the transverse momentum~\cite{css}, which has been accomplished 
to next-to-leading logarithmic order (NLL) for 
jet production processes~\cite{jetresum}, and up to third subleading order (N3LL) for vector 
boson~\cite{vresum} and 
Higgs boson production~\cite{hresum}. 
Through a dedicated matching procedure, these 
resummations are combined with the respective 
fixed-order results to yield predictions that 
are valid over the full kinematical range.

The vast majority of the measurements discussed in this review are unfolded for detector effects, which allows for a direct comparison with the state-of-the-art theoretical predictions, without any further need for using detector simulation.
They all strongly benefit from the excellent precision achieved in the detector performance studies at the LHC.
This includes the jet energy scale~(JES) and resolution~(JER) evaluations, based on detailed in-situ studies~\cite{ATLAS:2019oxp}, as well as the jet tagging where major improvements were achieved using jet substructure information and multivariate techniques~\cite{ATLAS:2019lwq}.

\section{QCD with hadronic final states}
\begin{figure}[t]
\parbox{0.58\columnwidth}{\includegraphics[width=0.56\columnwidth]{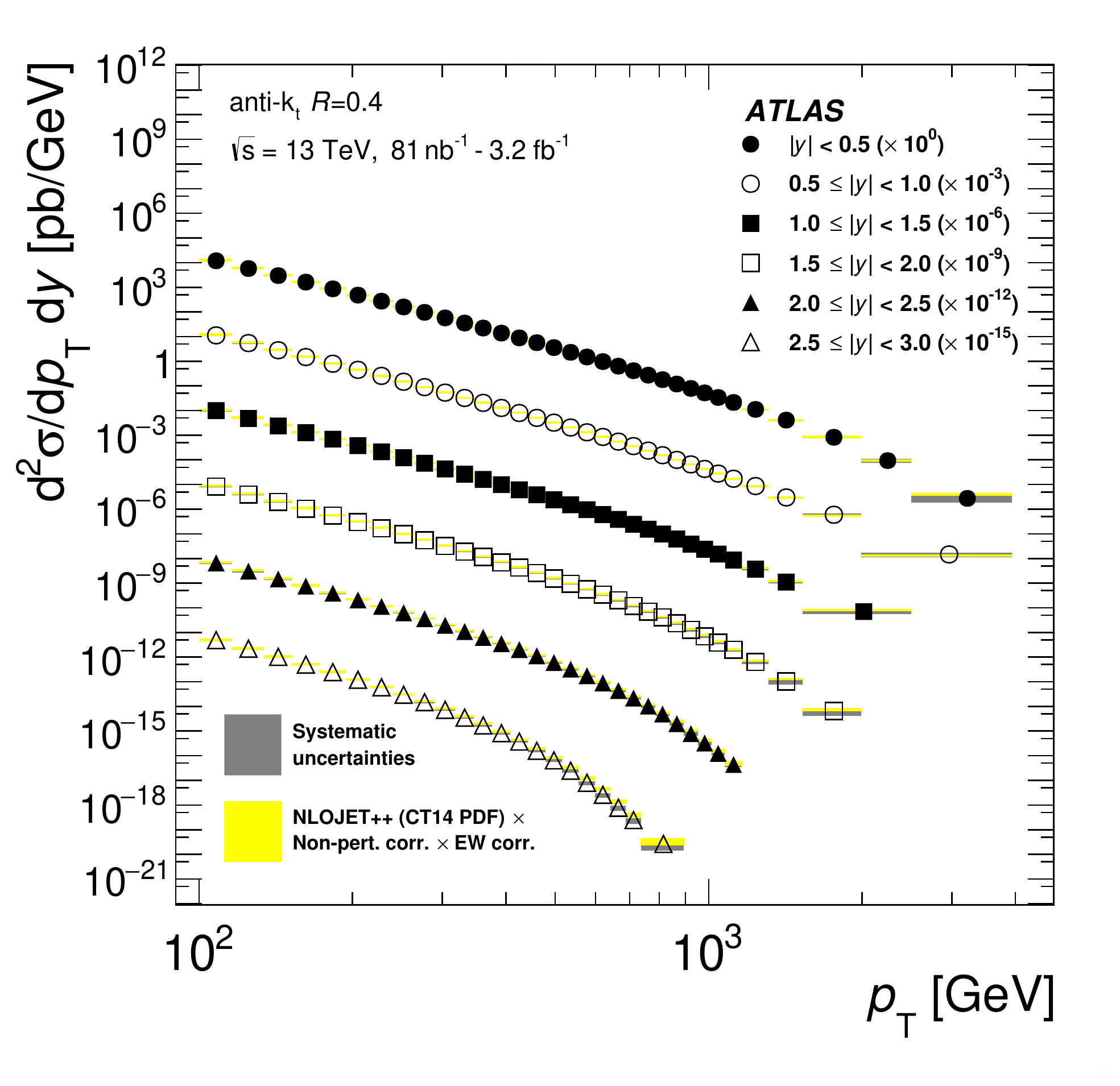}}
\parbox{0.4\columnwidth}{\includegraphics[width=0.4\columnwidth]{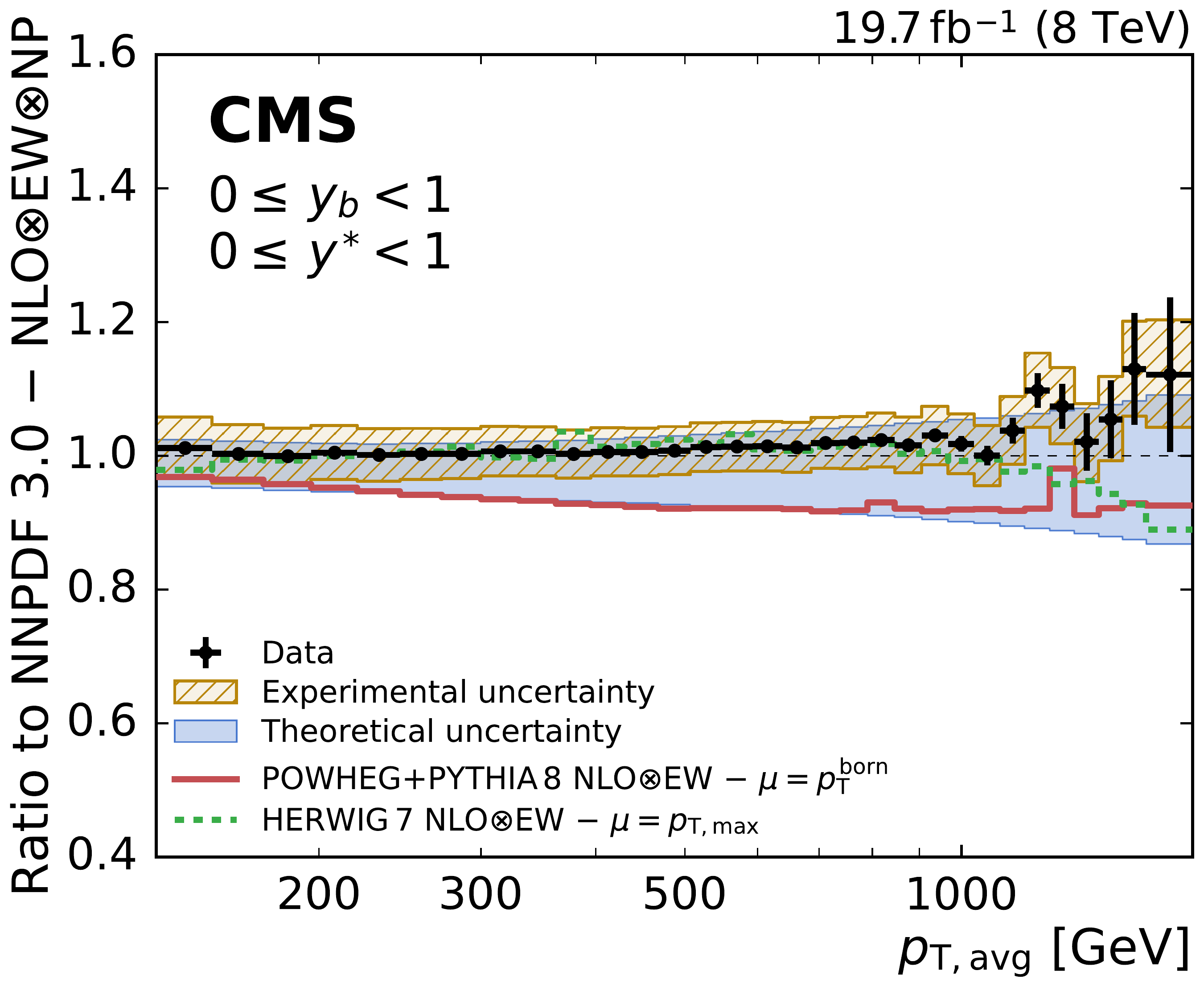}
\includegraphics[width=0.4\columnwidth]{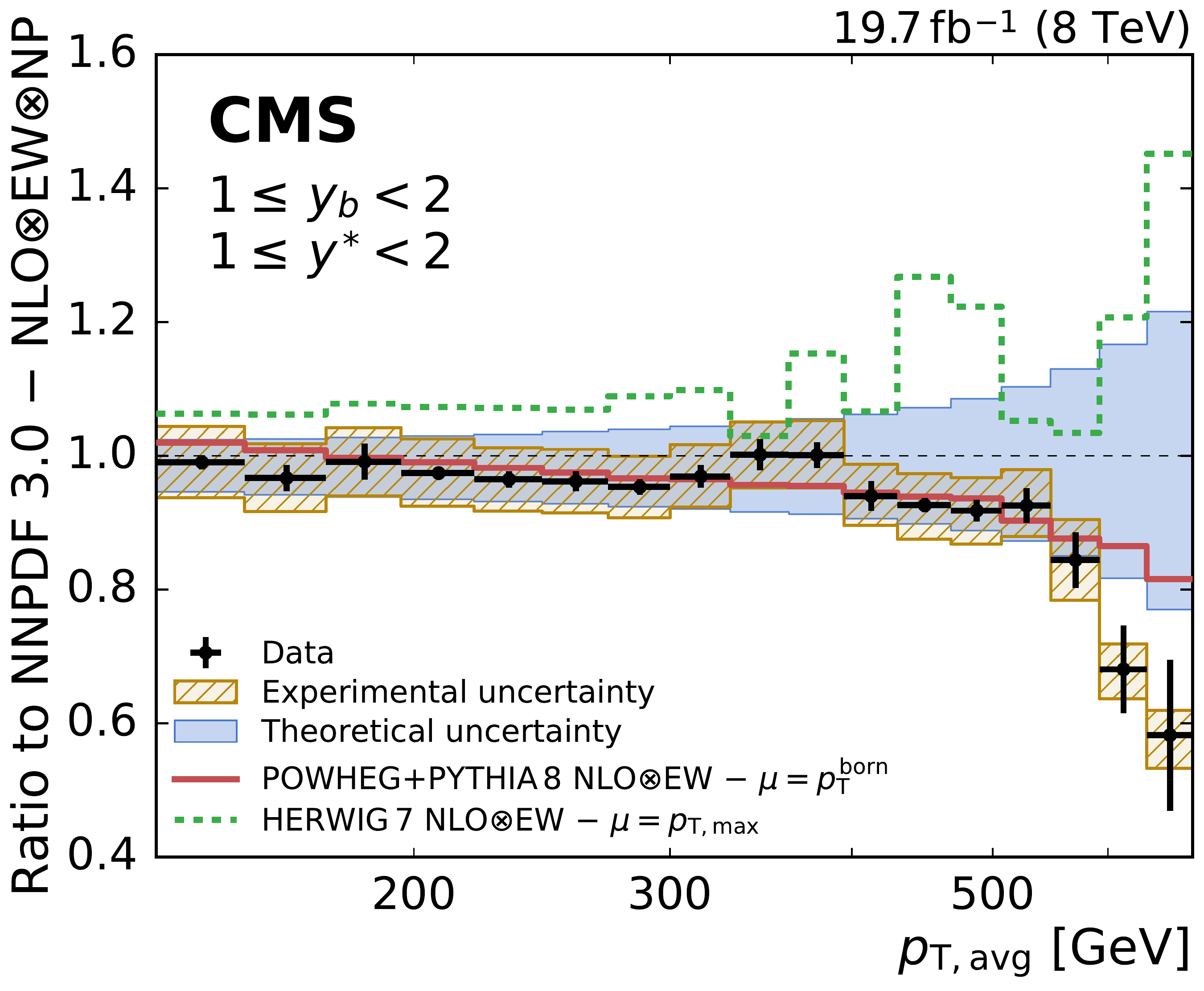}\\}
\caption{Left: ATLAS inclusive jet cross-sections as a function of $p_{T}$ and $|y|$, for anti-${\rm k}_{\rm t}$ jets with R=0.4, compared to NLO pQCD predictions calculated using NLOJET++~\cite{Nagy:2003tz} with $p_T^{max}$ as the QCD scale and the CT14 NLO PDF set, to which non-perturbative and electroweak corrections are applied. Figure from Ref.~\cite{ATLAS:2017ble}.
Right: Ratio of the triple-differential dijet cross section measured by CMS to the NLOJET++ prediction using the NNPDF3.0 set in the $(0 \leq y_b < 1, 0 \leq y^{\star} < 1)$ 
and the $(1 \leq y_b < 2, 1 \leq y^{\star} < 2)$ bins respectively. Figure from Ref.~\cite{CMS:2017jfq}.}
\label{fig1}
\end{figure}

Hadronic final states are produced most frequently at the LHC, and can be studied with very high statistical precision.
This is indeed the case for inclusive jet cross-sections~(see Refs.~\cite{ATLAS:2017ble,ATLAS:2017kux,CMS:2016jip,ALICE:2019qyj,CMS:2020caw} and references therein), for which unfolded multi-differential measurements have been performed~(see Figure \ref{fig1}).
Such measurements are dominated on a large fraction of the phase-space by JES and (to a somewhat lesser extent) JER uncertainties, while statistical uncertainties are relatively important only in the highest transverse momentum~($p_{\rm T}$) regions.
The definition of single-jet inclusive cross sections is peculiar, in the sense that all jets that are reconstructed in an event contribute 
individually to this distribution, in an additive manner and in different kinematical bins. 
This is in contrast to any other inclusive jet cross section (like inclusive di-jet production), where the reconstructed jets are ordered in 
their transverse momentum, then using only the leading jet(s) for the definition of the measured 
cross section. 
Even if the typical event selections yield an important overlap between the event samples used for inclusive jet and di-jet cross-section studies, the two observables probe the event topologies in complementary ways. 
While single-jet inclusive observables are sensitive to multiple radiation, the triple-differential measurements of di-jet final states~\cite{CMS:2017jfq} allow for a full reconstruction of the Born-level kinematics, thus providing direct PDF sensitivity. 
The statistical correlations caused by the overlap of the inclusive jet and di-jet samples can be quantified, as done in Ref.~\cite{ATLAS:2017ble}, using the bootstrap method described in Ref.~\cite{ATLAS:2013jmu}.

Figure \ref{fig1} compares the measured single-jet and di-jet inclusive cross sections to 
NLO QCD predictions~\cite{Nagy:2003tz}, augmented with NLO electroweak (EW) corrections and corrected for non-perturbative effects. 
While using the total experimental and theoretical uncertainties in each bin is sufficient for a qualitative comparison, the correlations between different phase-space regions play a key role for quantitative studies.
The impact of the fact that such correlations are not perfectly known, for both experimental and theoretical uncertainties, has been studied into some detail in Refs.~\cite{ATLAS:2017ble,ATLAS:2017kux}.
It is observed that the NLO QCD theory uncertainty 
on single-jet and di-jet production 
observables is typically larger than the total 
experimental uncertainty, thereby becoming the 
limiting element in quantitative comparisons. NNLO QCD predictions for  single-jet and di-jet production
were computed recently~\cite{our2j,czakon2j}. They lead to a substantial decrease of the theory uncertainty on these observables, improve the description of kinematical shape of the data, and enable the consistent inclusion of these data in PDF determinations at NNLO.  

Jet cross section measurements for varying values of jet resolution (R-scan) provide important information~\cite{ALICE:2019qyj,CMS:2020caw} on the formation dynamics of jets and validate the modelling of parton shower and hadronization models in multi-purpose event-generators. 
One of the main challenges for such measurements originates from the need to establish model-independent JES calibrations for every jet resolution value under consideration.
While deriving dedicated in-situ calibrations is indeed feasible for studies involving typically two jet resolution values~\cite{ATLAS:2017kux,CMS:2016jip}, in the existing R-scan studies one had to rely more on Monte Carlo simulation for the calibration.

The experimental reconstruction of charged and neutral hadrons is of different quality, and sometimes only charged particle momenta can be reconstructed with high resolution. Part of the physics content of jet cross sections can in this case be studied by charged-only jets (ALICE,~\cite{ALICE:2019wqv}) or 
through prompt charged-particle production (LHCb,~\cite{LHCb:2021abm}). 

With increasing jet multiplicity, novel types 
of observables can be constructed, thereby allowing 
to disentangle different dynamical effects. 
The most prominent example is the
determination of the strong coupling constant  $\alpha_{\rm s}$: single-jet or di-jet 
inclusive cross sections always probe the product of parton distributions and $\alpha_{\rm s}$, thus inducing a sensitivity on the external PDF input for the $\alpha_{\rm s}$ determination. 
In three-jet final states~\cite{ATLAS:2014qmg,CMS:2014mna}, the extra jet radiation beyond two-to-two scattering kinematics allows a more direct access to 
 $\alpha_{\rm s}$ from the three-jet cross section~\cite{CMS:2014mna}.
\begin{figure*}[t]
\centering
\parbox{0.4\columnwidth}{\includegraphics[width=0.4\columnwidth]{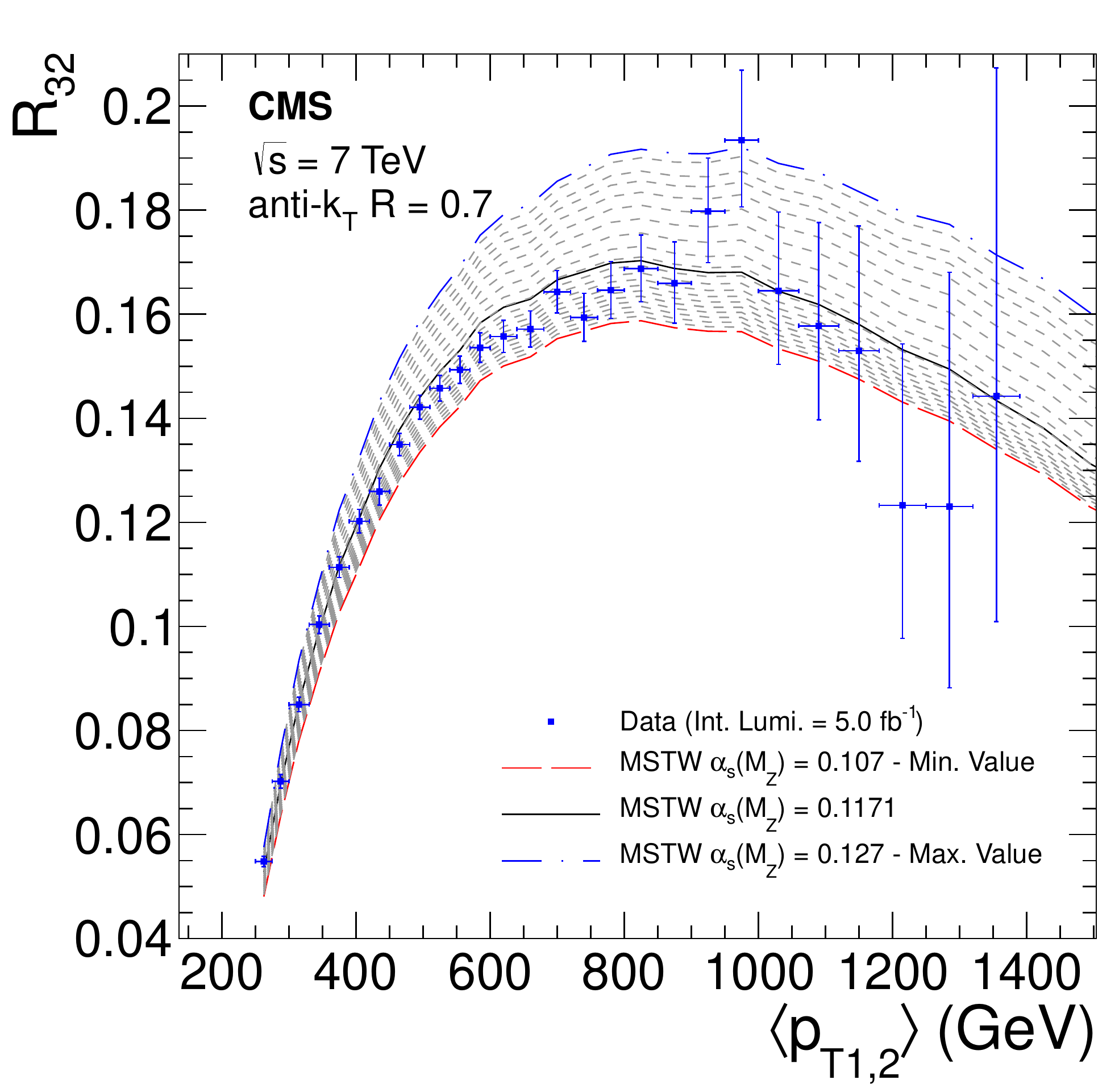}\\}
\parbox{0.5\columnwidth}{\includegraphics[width=0.5\columnwidth]{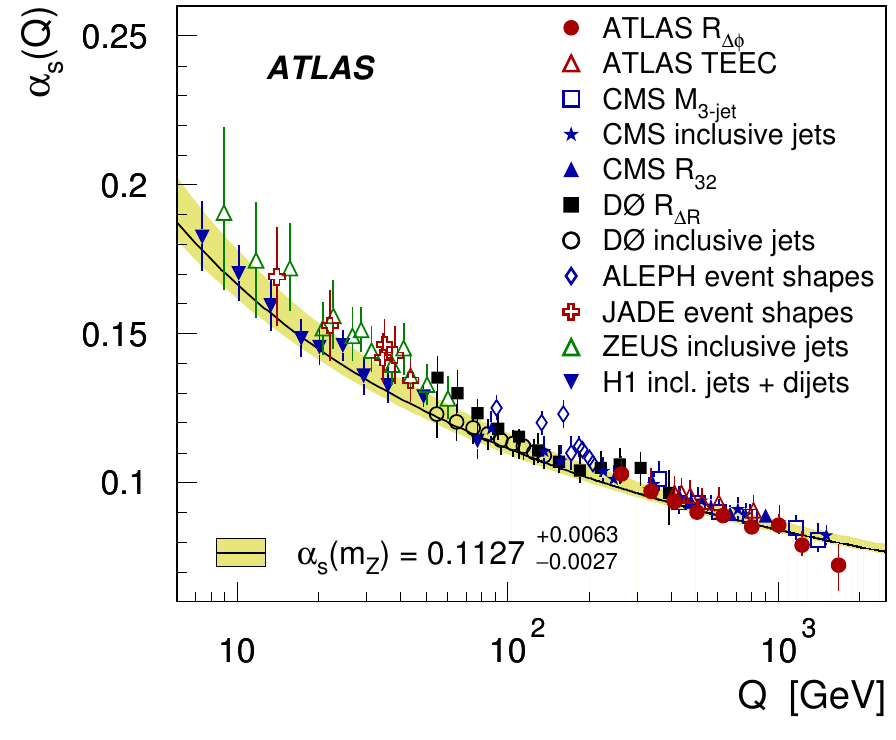}}
\caption{Left: The NLO predictions using the MSTW2008 NNLO PDF sets for a series of values of $\alpha_{\rm s}({\rm M}_{\rm Z})$ in the range $0.107-0.127$, in steps of 0.001, together with the R32 values measured by CMS. Figure from Ref.~\cite{CMS:2013vbb}.
Right: A selection of $\alpha_{\rm s}$ extractions from jet data. Figure from Ref.~\cite{ATLAS:2018sjf}.}
\label{fig3}
\end{figure*}

By constructing cross section ratios between three-jet and two-jet final states at fixed 
incoming parton kinematics, the PDF dependence can be reduced. The resulting $R_{32}$ ratio~\cite{CMS:2013vbb} is thus directly sensitive to $\alpha_{\rm s}$, see
Figure~\ref{fig3} (left). Three-parton final state kinematics can also be probed by 
the angular decorrelation $R_{\Delta\phi}$ in  two-jet final 
states~\cite{CMS:2016adr,ATLAS:2018sjf}, 
equally leading to an 
$\alpha_{\rm s}$ determination~\cite{ATLAS:2018sjf}. 
Up to now, $\alpha_{\rm s}$ determinations from three-jet-type observables were limited by the NLO QCD theory uncertainty. With 
NNLO corrections to three-jet production computed most recently~\cite{czakon3j}, this limitation will be overcome, thereby opening up these observables for precision QCD studies.  

A selection of collider determinations of $\alpha_{\rm s}$ is displayed in the right frame of Figure~\ref{fig3}, including 
three-jet mass~\cite{CMS:2014mna}, $R_{32}$~\cite{CMS:2013vbb}, $R_{\Delta\phi}$~\cite{ATLAS:2018sjf}, 
transverse energy-energy correlations~\cite{ATLAS:2017qir} (discussed below) 
as well as earlier measurements from LEP and HERA. 


For observables like $R_{32}$, $R_{\Delta\phi}$ and transverse energy-energy correlations, the scale used in the theoretical calculations is often chosen based on event-level quantities, like the average transverse momentum of the two leading jets~\cite{CMS:2013vbb,ATLAS:2017qir} or half of the scalar sum of the transverse momenta of all the selected jets in the event~\cite{ATLAS:2018sjf}.
Then, the evaluated $\alpha_{\rm s}$ values are displayed as a function of this same scale, reaching values up to a few TeV~(see Figure~\ref{fig3} right).
However, we note that for such observables the sensitivity to $\alpha_{\rm s}$ actually originates from the probability for emission of extra radiation, yielding a third or higher order jet.
Therefore, the energy scale at which these $\alpha_{\rm s}$ evaluations probe the prediction of the renormalisation group equation in QCD is rather related to the transverse momentum of the third jet~($p_{\rm T 3}$) than to the event-level quantities above.
It is to be noted that the typical values for $p_{\rm T 3}$ are significantly lower than the scale displayed in Figure~\ref{fig3} right~(for numerical values, see e.g.\ the appendix of Ref.~\cite{ATLAS:2017qir}).
The residual PDF uncertainty in the $\alpha_{\rm s}$ determinations from ratios of three-parton-like over two-parton-like final states has been quantified in the respective experimental papers~\cite{CMS:2014mna,CMS:2013vbb,ATLAS:2018sjf,ATLAS:2017qir}. Contrary to initial expectations it is non-negligible (typically larger than the combined experimental uncertainties, but smaller than the NLO scale uncertainty), possibly reflecting the different 
composition of partonic initial states in the three-parton and two-parton processes that are used to define the respective cross section 
ratios~\cite{Proceedings:2015eho}.

The azimuthal separation in nearly back-to-back jet topologies in inclusive two-jet and three-jet events~\cite{CMS:2019joc}
or in the presence of jet vetoes~\cite{ATLAS:2014lzu}
is sensitive to multiple soft radiation. They require resummation for a reliable theory description, which is also crucial for new 
physics searches in these types of topologies. 

Four-jet differential cross sections have been studied as a function of numerous observables~\cite{CMS:2017cfb,ATLAS:2015xtc,CMS:2016kdy}, providing detailed insights into event geometries and secondary production of flavour-tagged jets~\cite{CMS:2016kdy}.

Closely related to jet cross sections are 
distributions in event shape variables (ESV). These variables are constructed from the four-momenta of all particles in the event, and each variable expresses a certain geometrical feature of the event. A classical event shape variable is thrust, corresponding to the projection of all particle momenta on a principal axis of the event, which reaches its maximum for narrowly collimated two-jet final states and its minimum for spherically symmetric events. At hadron colliders, individual particle momenta cannot be reconstructed with sufficient quality, such that
event shape variables are typically constructed from jet momenta in multi-jet final states~\cite{ATLAS:2020vup,CMS:2018svp,CMS:2018vzn}. 
\begin{figure}[t]
\begin{center}\includegraphics[width=5in]{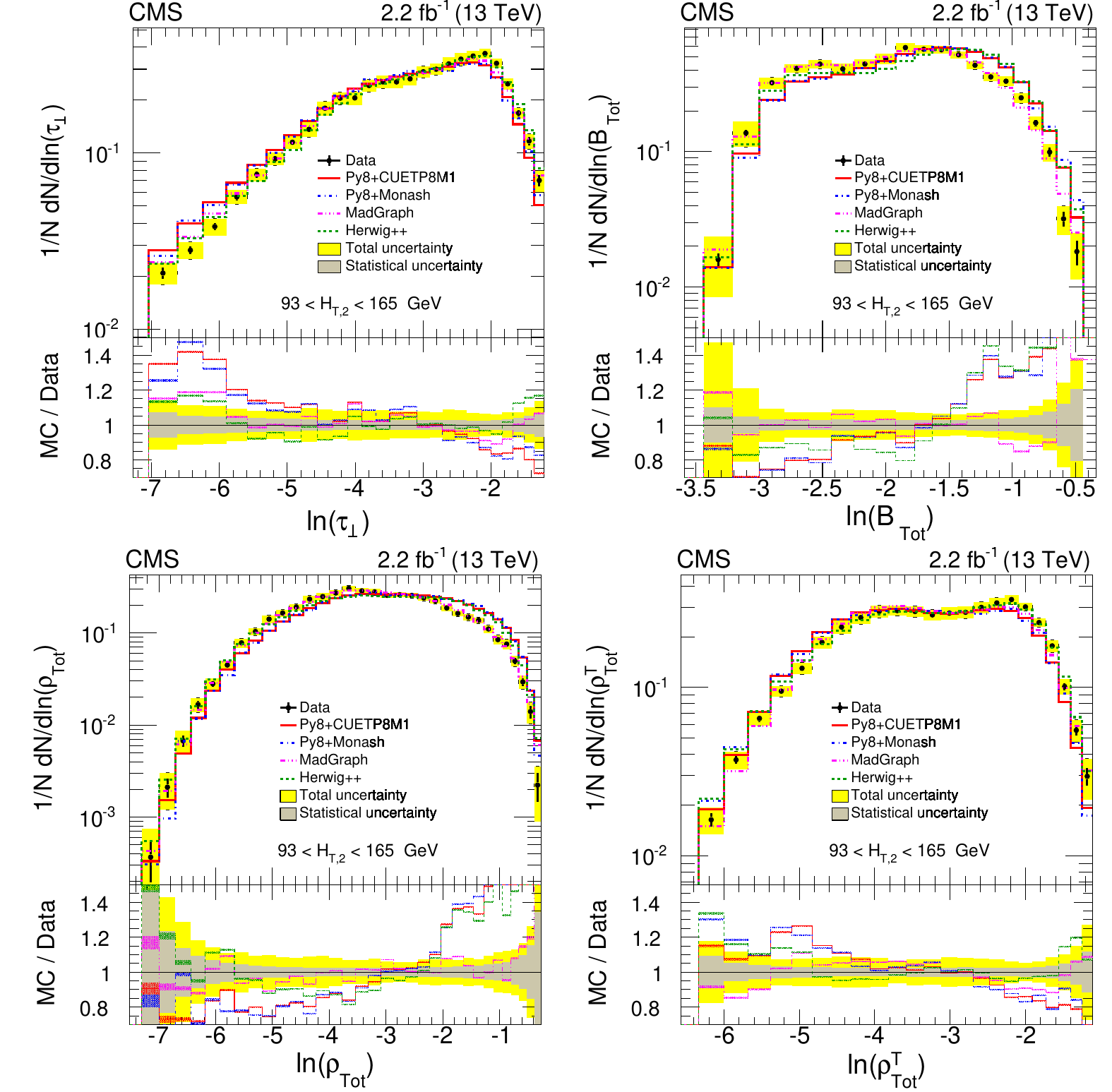}
\end{center}
\caption{CMS: Normalized differential distributions of unfolded data compared with theoretical Monte Carlo  predictions  as a function of ESV: complement of transverse thrust ($\tau_\perp$) (upper left), total jet broadening ($B_{{\rm Tot}}$) (upper right), total jet mass ($\rho_{{\rm Tot}}$) (lower left) and total transverse jet mass ($\rho^{T}_{{\rm Tot}}$) (lower right) all measured on events with $93 <H_{T,2}< 165$ GeV. See text for details, figure from Ref.~\cite{CMS:2018svp}.}
\label{fig2}
\end{figure}

Examples of event shapes measured by CMS~\cite{CMS:2018svp} are 
displayed in Figure~\ref{fig2}, showing the 
distributions in 
the complement of transverse thrust $\tau_\perp$, 
the total jet broadening $B_{{\rm Tot}}$, 
the total jet mass $\rho_{{\rm Tot}}$ and 
total transverse jet mass $\rho^{T}_{{\rm Tot}}$. 
In each ratio plot, the inner gray band represents statistical uncertainty and the yellow band represents the total uncertainty (systematic and statistical components added in quadrature) on data and the MC predictions include only statistical uncertainty.
All variables are defined in a way that the two-jet limit corresponds to the left edge of the plot and the multi-jet limit to the right edge. 
The data are compared to the Monte Carlo event generators PYTHIA~\cite{pythia} (for two different parameter tunes), HERWIG++~\cite{herwig} and mg5\_aMCatNLO~\cite{mg5}. Substantial differences between the generators are observed in the two-jet limit, exposing the differences in parton shower prescriptions and hadronization models. Towards the multi-jet limit, mg5\_aMCatNLO in general yields the better description, reflecting its fixed-order NLO 
accuracy (matched for different multiplicities) over the LO accuacy of the other programs. 
Event shape distributions receive non-trivial 
contributions only from final states with at least three partons (the two-parton contribution corresponding to a fixed end-point of the distribution), such that they display a similar sensitivity to $\alpha_{\rm s}$
as three-jet production. Likewise, their fixed-order QCD computation is analogous to three-jet production. 

Another class of event shape variables are 
energy-energy correlators (EEC), 
which are conceptually 
different from the ESV considered above. While thrust, broadening and jet masses assign each event a unique value of the ESV, which is based on 
including all object (particle or jet) momenta in the event, EECs are computed on all individual object pairs in the event, such that each event provides multiple contributions  to an EEC distribution (as in the case of the single-jet inclusive cross section). 

Transverse energy-energy correlations (TEEC) and 
associated asymmetries were measured by ATLAS and 
used in a determination of $\alpha_{\rm s}$~\cite{ATLAS:2017qir}, see also Figure~\ref{fig3} (right). EECs for two and more objects
are particularly attractive observables since analytical predictions for their 
distributions can be computed from first 
principles~\cite{Dixon:2019uzg}, thereby probing fundamental symmetries of QCD. 
They also form the basis of novel geometrical measures of event properties~\cite{Komiske:2020qhg} that go beyond jet definitions and ESVs.

Jet algorithms typically combine individual particle momenta into jet momenta in an iterative procedure based on distance measures, aiming to reconstruct 
the momentum of a primary parton produced in a hard scattering interaction or in a heavy particle decay. They potentially attribute unrelated particles, 
which can be from an underlying event, from multiple interactions, from soft exchanges and hadronization or from overlap with neighbouring objects, to the jet. Various techniques (e.g.\ 
trimming, pruning, grooming) 
have been proposed to improve the kinematical reconstruction of jets. These techniques also 
enable the resolution of jet substructure~\cite{Kogler:2018hem}, especially 
in view of separating hard and soft dynamics, and 
to reconstruct decay products from boosted 
heavy particles (e.g.\ weak gauge bosons decaying hadronically) inside a jet. 
A particularly powerful substructure observable is 
soft drop, which acts on the full clustering 
history of each jet, and
grooms the jet by removing branches corresponding to large-angle soft clusterings.  Soft-drop can not be applied directly on the  anti-$k_T$ jet algorithm that is the LHC default, but requires the jet clustering history to be re-derived in an angular ordered algorithm (Cambridge-Aachen or $k_T$). 
The resulting cross sections
measured on soft-drop groomed jets~\cite{ATLAS:2017zda} are less sensitive to hadronization corrections, and several of the observables derived from them (e.g.\ jet masses) 
can be related more directly to theory expectations. 
\begin{figure}[t]
\begin{center} \parbox{0.4\columnwidth}{\includegraphics[width=0.4\columnwidth]{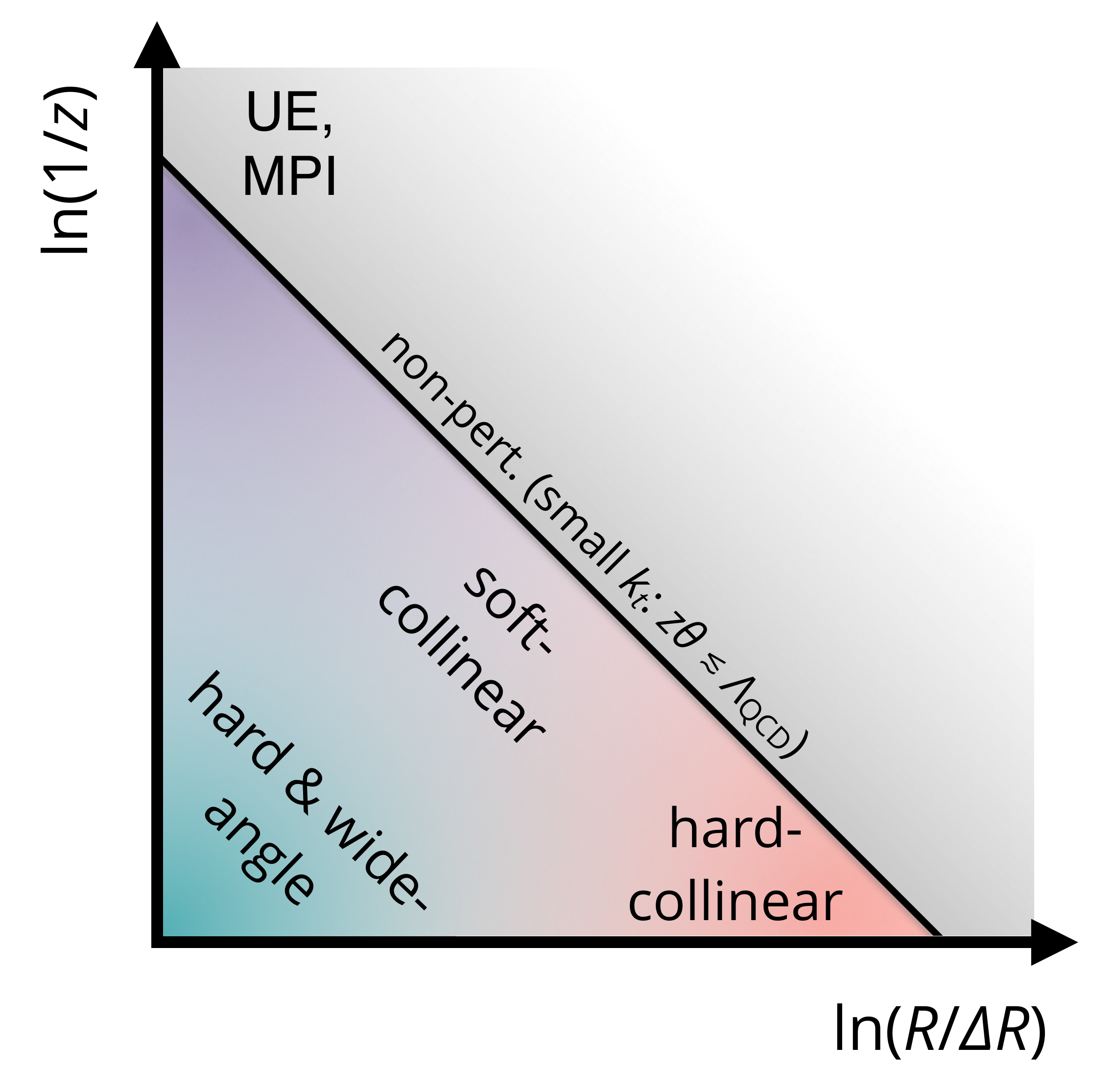}}\hspace{-0.1cm}
\parbox{0.4\columnwidth}{\includegraphics[width=0.4\columnwidth]{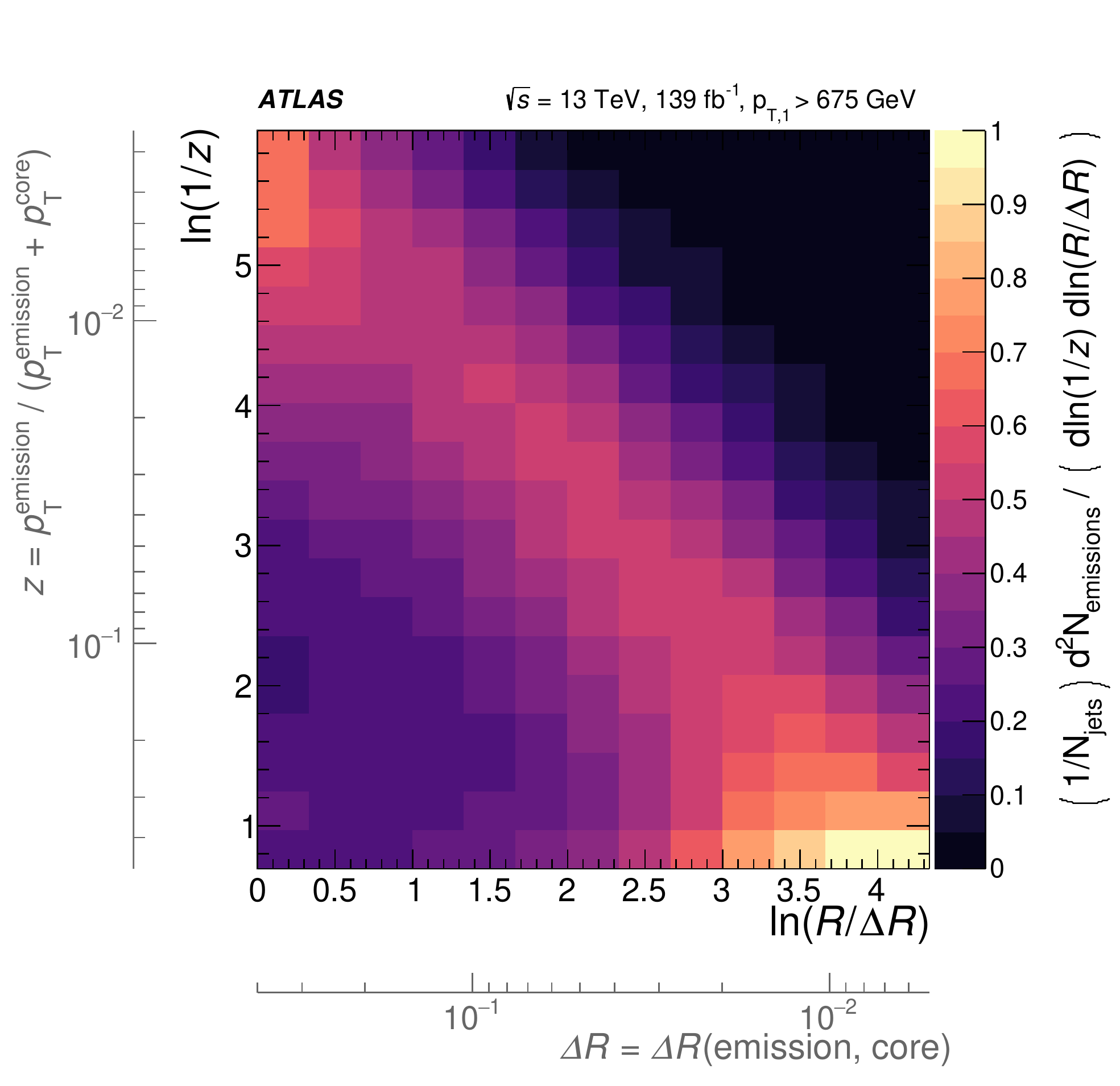}}\\
\parbox{0.4\columnwidth}{\includegraphics[width=0.4\columnwidth]{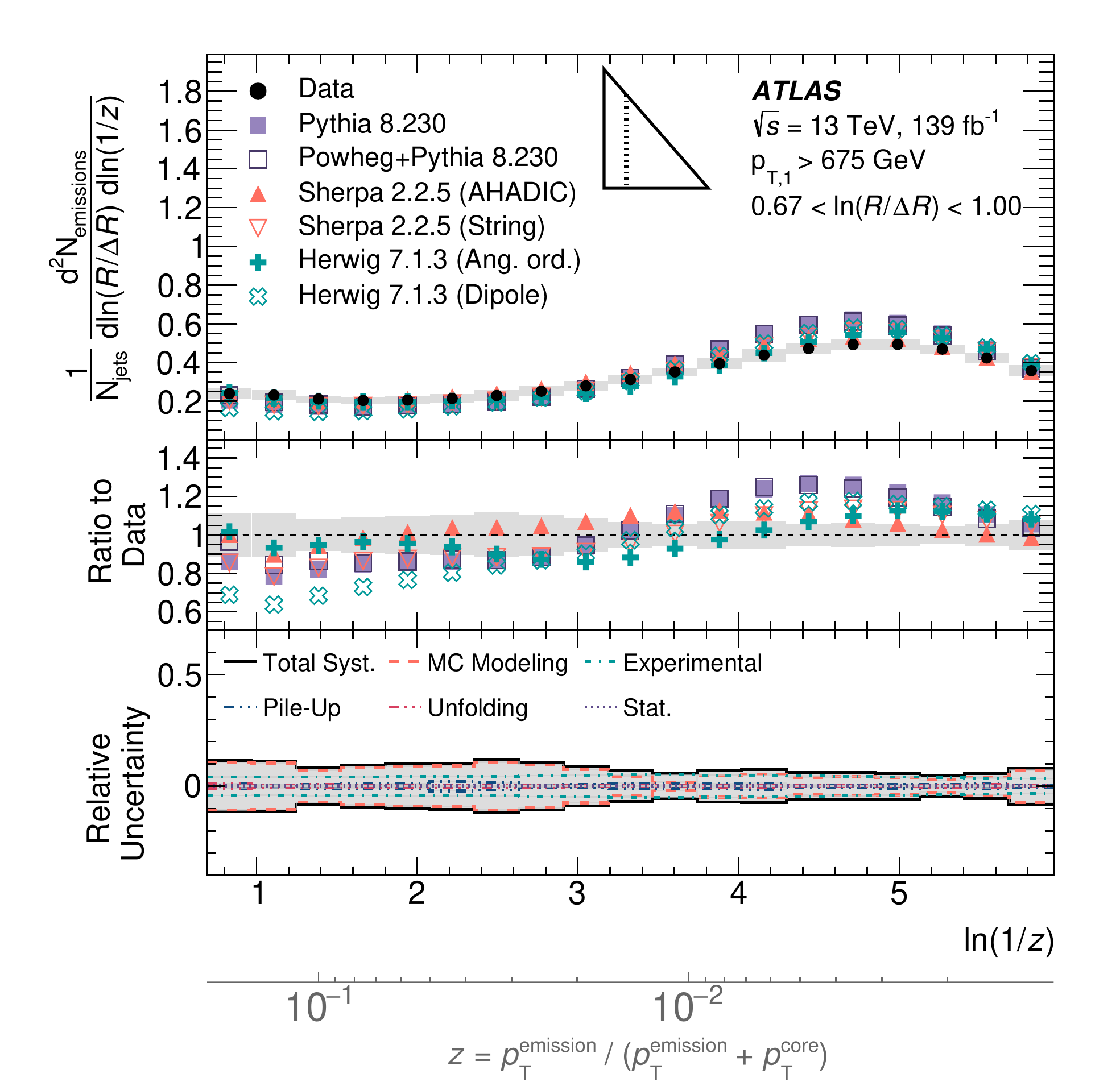}\\}
\parbox{0.4\columnwidth}{\includegraphics[width=0.4\columnwidth]{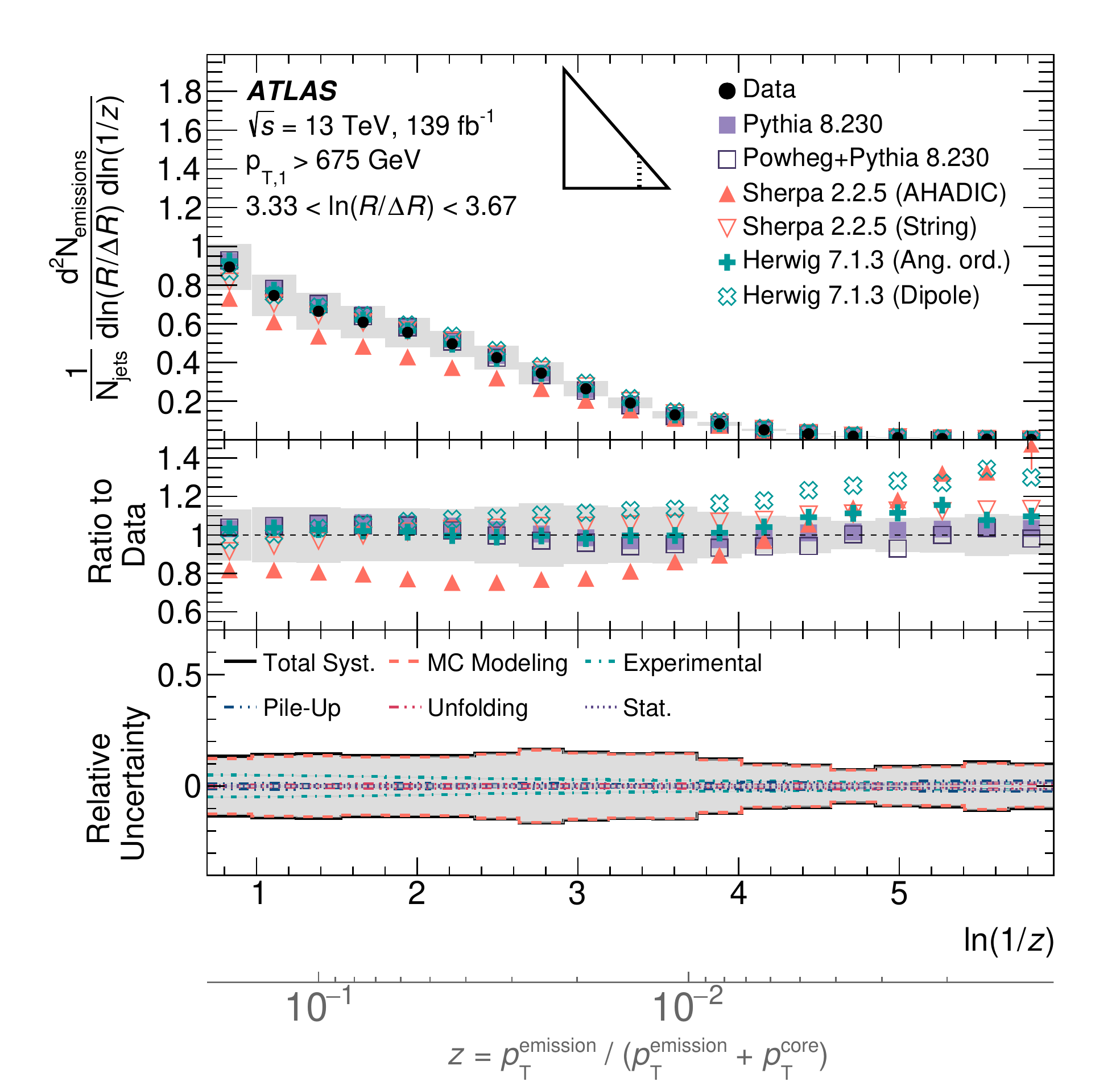}\\} 
\end{center}
\caption{(a) Schematic representation of the LJP. 
(b) The LJP measured by ATLAS using jets in 13 TeV data, corrected to particle level. 
(c,d) Representative vertical slices through the LJP. Unfolded data are compared with particle-level simulation from several MC generators. See text for details,
figures from Ref.~\cite{ATLAS:2020bbn}.}
\label{fig4}
\end{figure}

The Lund Jet Plane, LJP,~\cite{Dreyer:2018nbf} provides 
a two-dimensional representation of the internal structure of a jet, in terms of the transverse momentum $k_T$ and the opening angle $\theta$ of a particle emission with respect to its emitter. In experimental measurements~\cite{ATLAS:2020bbn,ALICE:2021Lund}, these variables are very hard to assess at the level of individual splittings. They are therefore approximated by the fraction $z$ of transverse momentum (relative to the beam direction) and the clustering distance 
$\Delta R$ of the particle emission with respect to the emitter. A full clustering history of a single event produces many entries into the Lund plane, and a full event sample yields a density distribution over the Lund plane. 
The LJP plane allows for a clear separation of 
different perturbative and non-perturbative dynamics that induce parton splittings and that populate the particle distribution inside a jet. Hard wide-angle emissions 
correspond to large $k_T$ (i.e.\ large $z$) and large opening angle $\theta$ (large $\Delta R)$, hard-collinear emissions have large $z$ at small $\Delta R$ while soft emissions are at small $z$ and any $\Delta R$. 
The boundary to non-perturbative dynamics is reached 
when $k_T$ becomes of order $\Lambda_{\rm QCD}$, 
translating to a line in a logarithmic ($z,\Delta R$) 
representation. Underlying event (UE) and multi-particle interactions (MPI) contribute in the non-perturbative domain at very small $k_T$ and large angles.
A schematic graph of the LJP is 
given in the upper left frame of Figure~\ref{fig4}, indicating the different types of parton splittings corresponding to different regions. 
The upper right frame of  
Figure~\ref{fig4} displays the 
 ATLAS measurement~\cite{ATLAS:2020bbn} 
 of the LJP density using jets in 13 TeV LHC data, corrected to particle level. 
The inner set of axes indicates the coordinates of the LJP itself, while the outer set indicates corresponding values of $z$ and $\Delta {\rm R}$.
The two lower frames show
representative vertical slices through the LJP, and unfolded data are compared with particle-level simulation from several MC generators. The uncertainty band includes all sources of systematic and statistical uncertainty. The inset triangle illustrates which slice of the plane is depicted: (lower left) $0.67 < {\rm ln}({\rm R}/\Delta {\rm R}) < 1.00$, (lower right) $3.33 < {\rm ln}({\rm R}/\Delta {\rm R}) < 3.67$.
The particle-level simulations provide a good qualitative description of the data. At the quantitative level, differences are observed especially among different parton shower prescriptions. Angular-ordered showers are in general providing a better description of the observed data, although none of the shower prescriptions agrees with the data over the full phase space. 

Jet cross sections, event shape variables and 
Lund jet plane distributions are insensitive to the hadron species that enter their reconstruction. 
Through the identification of specific hadrons inside jets, 
additional information on the jet production dynamics can be gained. Most prominent in this context are heavy-flavour (charm or bottom) jets, that are identified through the presence of specific heavy hadrons or by the reconstruction of a secondary interaction vertex. For example,
LHCb has measured $ b\overline{b} $- and $ c\overline{c} $-dijet cross-sections in the forward region~\cite{LHCb:2020frr}, charm jets 
were tagged with  D$^{0}$ mesons by ALICE~\cite{ALICE:2019cbr}, and 
ATLAS investigated  
$g\rightarrow b\bar{b}$ 
splitting~\cite{ATLAS:2018zhf}. 
Light-flavoured hadrons are produced copiously in QCD jets. Their prevalence can only be studied in a statistical manner, and their production is modelled in event simulations programs through hadronization models whose parameters are tuned to experimental data.  Various studies have examined the 
jet fragmentation into hadrons~\cite{ATLAS:2019rqw}, charged hadron production in $Z$-tagged jets~\cite{LHCb:2019qoc} and 
the net jet charge in di-jet events~\cite{ATLAS:2015rlw}.
Through comparisons with event simulations, these properties can be loosely correlated with the parton species that induced the jet. Recent advances in 
the application of machine learning 
techniques~\cite{Komiske:2016rsd}
 hold the promise of a far 
more efficient separation of 
quark jets and gluon jets.

Differential cross sections in top quark pair production~\cite{ATLAS:2019hxz,CMS:2021vhb}
 provide phenomenological information on the top quark production dynamics, thereby testing the Standard Model in the top sector. They also 
 yield complementary information on the gluon PDF, 
 with recent advances on NNLO theory predictions~\cite{czakon1,grazzinitop} enabling their consistent inclusion in PDF fits at this order. While top quark measurements are in general performed by lepton+jets final states, which are easier to identify, the all-hadronic top quark 
decay channels are now beginning to reach a 
competitive precision level for differential 
measurements~\cite{ATLAS:2020ccu} and for highly boosted top quarks~\cite{ATLAS:2018orx}. 
More information on the
top quark production dynamics be can 
gained from 
${t} \bar{t} + {\rm jets}$ production~\cite{ATLAS:2018acq}, 
possibly also enabling  non-kinematical 
measurements of the top quark mass~\cite{ATLAS:2019guf}.


\section{QCD in vector-boson (gamma,W,Z) final states }

 Final-state vector bosons ($\gamma$, W, Z) provide very clean experimental signatures. The neutral vector bosons  also offer excellent kinematical reconstruction. 
 Compared to Z boson production, observables 
that involve W bosons have larger 
production cross sections. They are nevertheless 
more challenging to measure to high precision, 
due to their incomplete kinematical reconstruction 
with the final-state neutrino only being inferred from missing transverse momentum (MET).  The vector boson production dynamics is covered by QCD processes, and they provide an excellent ground for precision studies. Fully inclusive vector boson production is studied as a function of rapidity and transverse momentum. For W and Z production, their decay products are detected only in a finite acceptance range, such that these measurements are only performed for a predetermined set of fiducial cuts on the leptons and missing energy, which need to be accounted for when comparing to theory. In turn, the spatial distribution of the decay products provides extra information in the form of angular coefficents or cross section 
 asymmetries that connect the 
 massive vector boson production and decay dynamics.
 More exclusive information is gained from vector boson production in association with jets.  

Among the vector boson production processes, photons have the largest production cross sections. 
Since photons can also be produced as decay products of hadrons, the measurement of cross sections involving final-state photons requires a photon isolation procedure to single out photons produced directly in the hard interaction process. 
This is achieved
in the experimental measurements by allowing 
only a limited amount of hadronic energy in a cone 
around the photon direction, and challenges the theory predictions to reproduce this isolation prescription as closely as possible.  

Cross sections for isolated photon and 
photon-plus-jet production have been measured by 
ATLAS~\cite{ATLAS:2017xqp} and CMS~\cite{CMS:2018qao} reaching a precision of a few per-cent in wide kinematical regions, and 
requiring theory predictions to NNLO accuracy~\cite{gamj} to match the precision of the data. 
Two-photon final states  
are studied frequently in Higgs precision measurements, and multi-photon production is an important signature in new physics searches. Cross section measurements 
of the non-resonant production are available 
from ATLAS
for two-photon~\cite{ATLAS:2021mbt} and three-photon 
final states~\cite{ATLAS:2017lpx}. 
\begin{figure}[t]
\parbox{0.51\columnwidth}{\includegraphics[width=0.51\columnwidth]{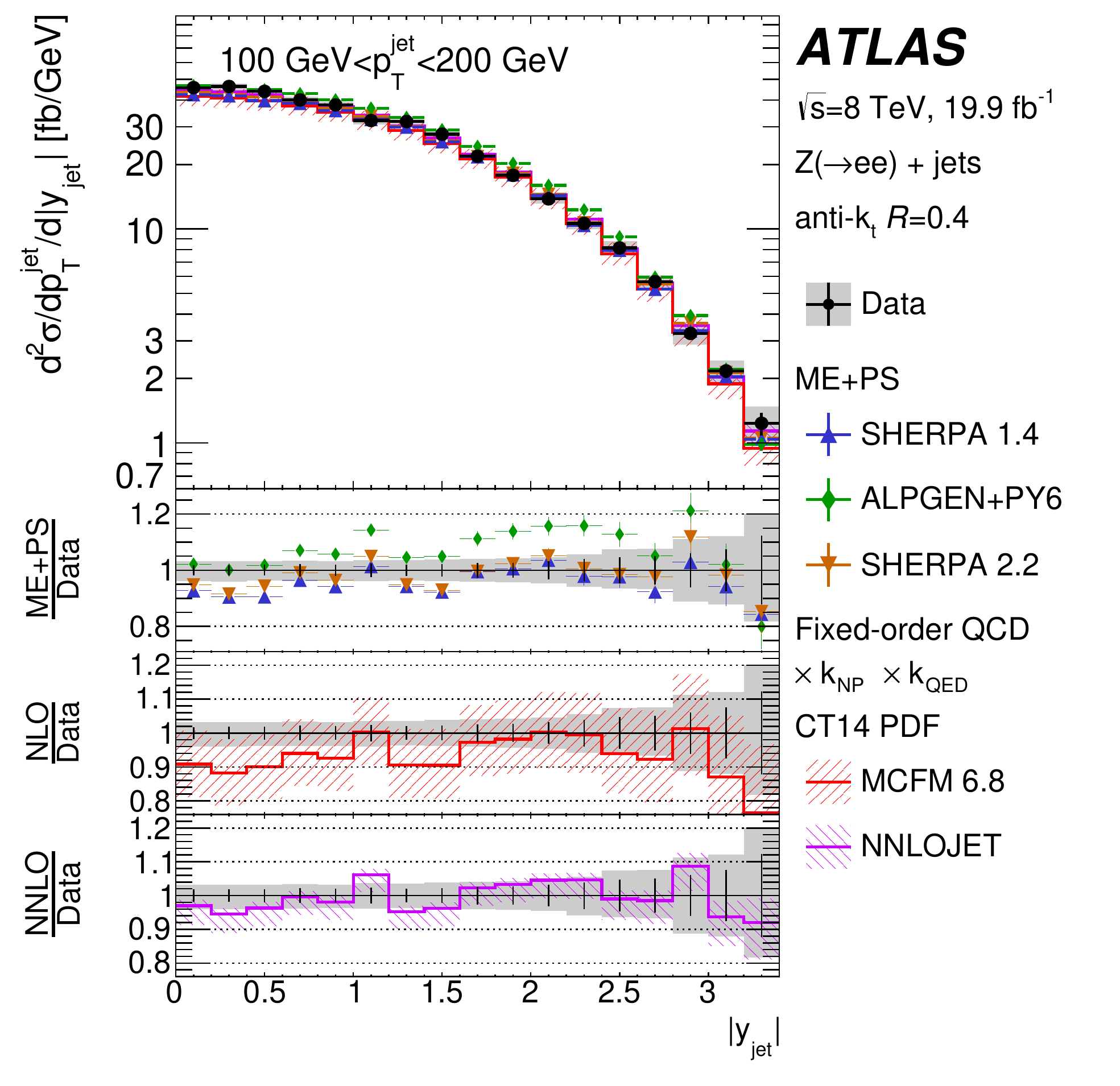}}\hspace{-0.01\columnwidth}
\parbox{0.36\columnwidth}{\includegraphics[width=0.36\columnwidth]{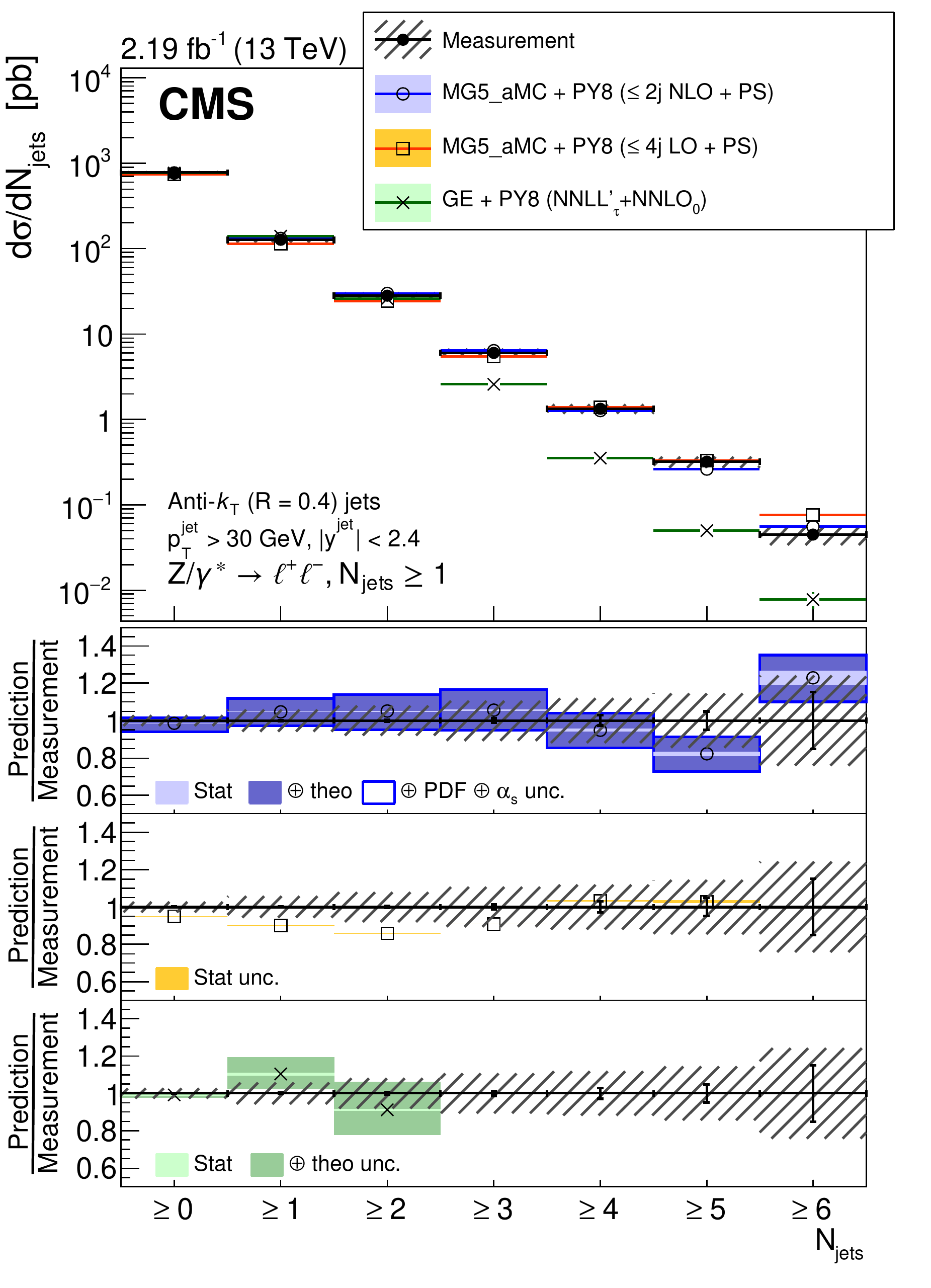}}\\
\parbox{0.9\columnwidth}{\includegraphics[width=0.9\columnwidth]{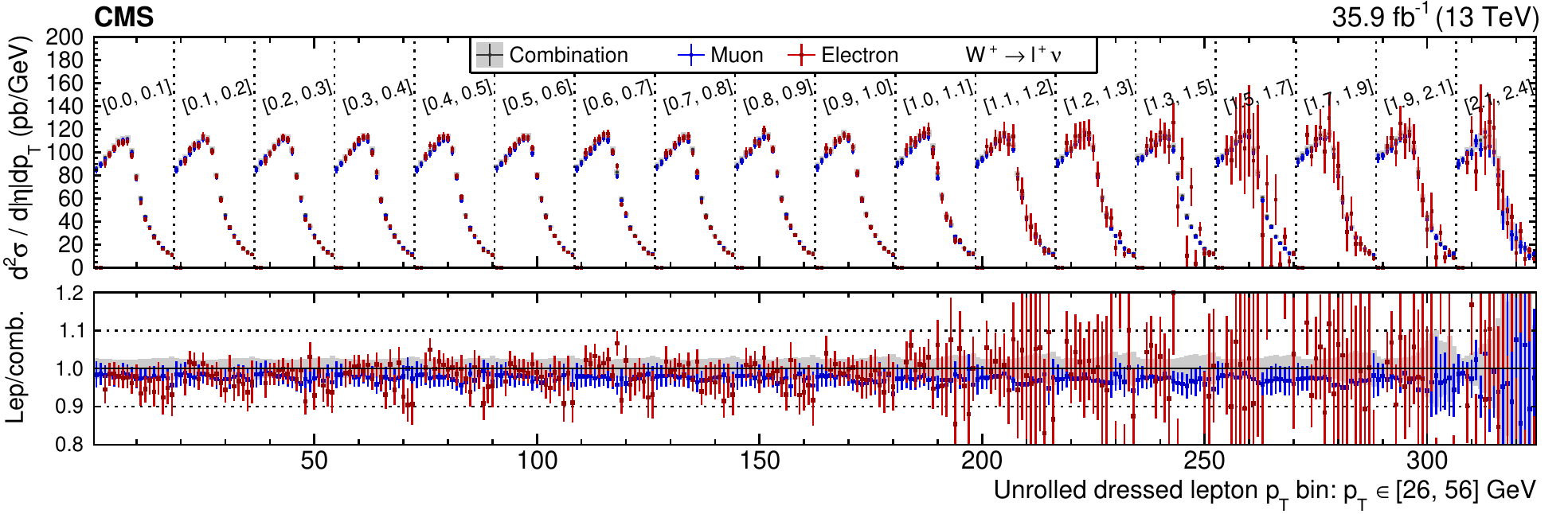}\\}
\caption{Top left: Double-differential Z+jet cross-section measured by ATLAS, as a function of $|y_{\rm jet}|$ in the $100~{\rm GeV} < p_{\rm T}^{\rm jet} < 200~{\rm GeV}$ range. See text for details, 
Figure taken from~\cite{ATLAS:2019bsa}. Top right:  
Cross section for Z+jets measured by CMS, as a function of the jet inclusive multiplicity. See text for details, Figure taken from~\cite{CMS:2018mdf}.
Bottom: Absolute double-differential $W^{+}$ boson production cross section measured by CMS, as function of 
the charged lepton variables 
$p_{\rm T}^{\rm l}$ and
$|\eta^{\rm l}|$. See text for details, Figure from~\cite{CMS:2020cph}.
}
\label{fig5}
\end{figure}

Differential cross sections in Z+jets final states have been measured by ATLAS~\cite{ATLAS:2019bsa}, CMS~\cite{CMS:2018mdf} and LHCb~\cite{LHCb:2016nhs} over a broad kinematical range and for various LHC energies: the ATLAS and CMS data at central rapidities are complemented by LHCb at forward rapidities. 
Two examples are shown in the upper frames of Figure~\ref{fig5}. 

An example for 
 double-differential Z+jets cross-sections measured by ATLAS~\cite{ATLAS:2019bsa} using its 8 TeV data set 
 is shown in the top left frame of Figure~\ref{fig5}, as a function of $|y_{\rm jet}|$ in the $100~{\rm GeV} < p_{\rm T}^{\rm jet} < 200~{\rm GeV}$ range. 
The statistical uncertainties are shown with error bars. The total uncertainties in the measurement~(summing the statistical and systematic uncertainties, except for the luminosity uncertainty of $1.9\%$) and in the fixed-order theory predictions~(summing in quadrature the PDF, scale, $\alpha_{\rm s}$, non-perturbative and QED radiation correction uncertainties) are represented with shaded bands. 
At central jet rapidity, the total measurement uncertainty is below 5\%, it increases only gradually towards forward rapidities, exceeding 5\% at 
$|y_{\rm jet}|=2$. 
The data are compared with different parton shower MC generator predictions and with the fixed-order theory predictions corrected for non-perturbative and QED radiation effects. Lower panels show the ratios of predictions to data. 
It is observed that the parton shower MC programs provide a good description of the overall features of the data, predictions from different programs are however scattered in an interval of about 10\%. This scatter is due to different parton shower models and 
different technical choices in the MC programs. It is very difficult to quantify the uncertainty on the MC predictions in quantitative terms, which is why no uncertainty bands are given on them.
For the fixed-order theory predictions at NLO~\cite{mcfm}, the estimated uncertainty is above 10\%, which is larger than the measurement uncertainty and thus limiting comparisons at precision level. Inclusion of NNLO corrections~\cite{ourvj} leads to a substantial decrease of the theory uncertainty to a level of about 4\% at central rapidity and provides a better description of the shape of the data than MC parton shower or NLO predictions. 

An example of a Z+jets cross section measurement by CMS~\cite{CMS:2018mdf}
is shown in the top right frame of Figure~\ref{fig5}, 
 as a function of the inclusive jet multiplicity. 
The error bars represent the statistical uncertainty and the grey hatched bands represent the total~(systematic and statistical) uncertainty. The measurement is compared with different predictions, based on the combination of fixed-order LO or NLO predictions with parton showers, MG5\_aMCatNLO~\cite{mg5} and with an NNLO calculation of Z+0jet final states matched onto resummation~\cite{Alioli:2015toa}, which yields at most final states with two jets.  
The ratio of each prediction to the measurement is shown together with the prediction uncertainties~(coloured bands), with a similar decomposition as above. The statistical~(black bars) and total~(black hatched bands) uncertainties are also displayed. All three theory calculations provide a very good agreement with the data where applicable, demonstrating that the extra jet radiation is well-described.
Only the NLO+PS calculation
yields predictions with uncertainties (which is, however, quantified on the fixed-order part only and does not include parton shower uncertainties) that match the precision of the data.

Collinear Z boson production and 
cross section ratios for 
$\gamma$ + jets and Z+jets production at fixed kinematics have been measured by CMS~\cite{CMS:2021fxy}. These ratios are an important ingredient to data-driven background estimates~\cite{Lindert:2017olm}, for example in mono-jet searches (see Section~\ref{sec:searches} below), by using the larger production cross sections for photons to predict cross sections for Z bosons (decaying to neutrinos), which have similar coupling structure and partonic composition. 

Ratio distributions between three-jet production and Z-plus-two-jet production are very sensitive to the interplay of soft and hard QCD radiation. By investigating these cross sections ratios as function of jet transverse momenta and opening angles, CMS~\cite{CMS:2021hnp} observed that parton-shower event generators provide a very good description of the data in regions that are dominated by soft dynamics, while the behaviour in harder regions (e.g.\ at large opening angles) is accounted for only if the parton shower is supplemented by higher-order matrix element calculations.

 Vector boson pair production is studied primarily as a probe of the coupling structure of the electroweak interaction. 
 Vector boson scattering (VBS), 
 which yields a final state with a vector boson pair and two  forward jets is 
 particularly relevant due to its sensitivity to the interplay of gauge and Higgs interactions. The VBS process cannot be strictly separated from other processes yielding identical final states, but kinematical selection criteria can help to enhance its contribution. For this, an in-depth understanding of the QCD dynamics in events with vector boson pairs and jets is crucial, and first measurements in this direction are being undertaken~\cite{CMS:2018ccg}. 
 The enhanced statistics collected during the high luminosity phase of the LHC will be greatly beneficial for the study of such rare processes. For many of them it will allow to move from the observation phase to precise (differential) cross-section measurements, as well as to study the different polarisation states, probing hence the mechanism of the electroweak symmetry breaking.

The bottom panel of figure~\ref{fig5} displays an 
example of double-differential 
absolute cross sections in W$^+$ production, 
measured by CMS~\cite{CMS:2020cph}. The 
data are unrolled in a one-dimensional histogram in 
the charged lepton transverse momentum 
$p_{\rm T}^{\rm l}$ in bins of the lepton 
rapidity $|\eta^{\rm l}|$.
 The markers indicate the muon-channel~(blue), electron-channel~(red) and combined~(gray) fits. The error bars indicate the total uncertainty from the respective fits. The filled gray band indicates the total uncertainty from the combined fit.
 Such multi-differential measurements~\cite{CMS:2020cph}
 and related cross section ratios between W$^+$ and 
 W$^-$ production~\cite{ATLAS:2017irc}
 can then be used to determine charge asymmetries that are relevant to quark PDF studies, and 
 provide probes of the chirality structure of the 
 W boson couplings. 

 \begin{figure}[t]
\parbox{0.5\columnwidth}{\includegraphics[width=0.5\columnwidth]{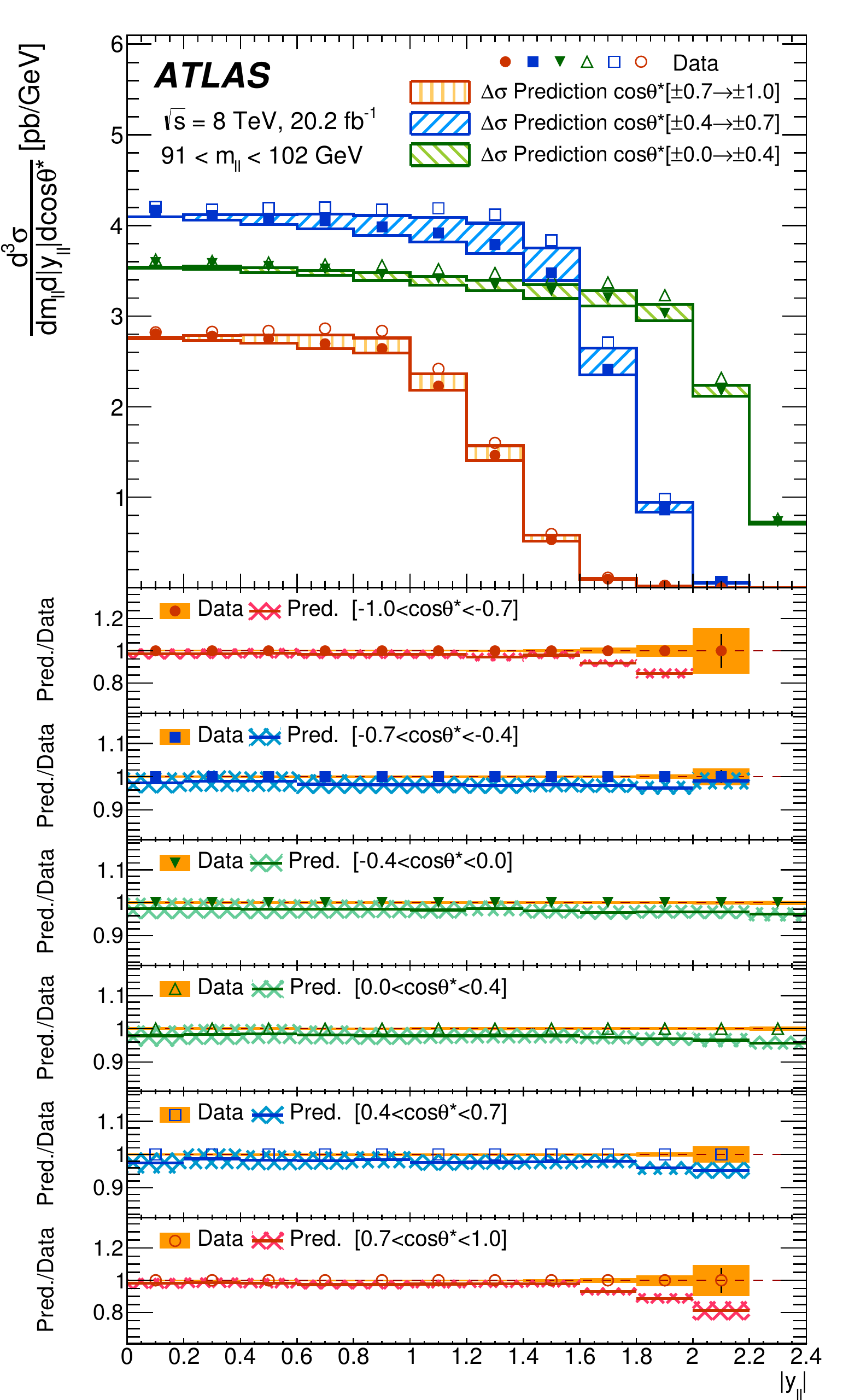}\\}
\parbox{0.5\columnwidth}{\includegraphics[width=0.5\columnwidth]{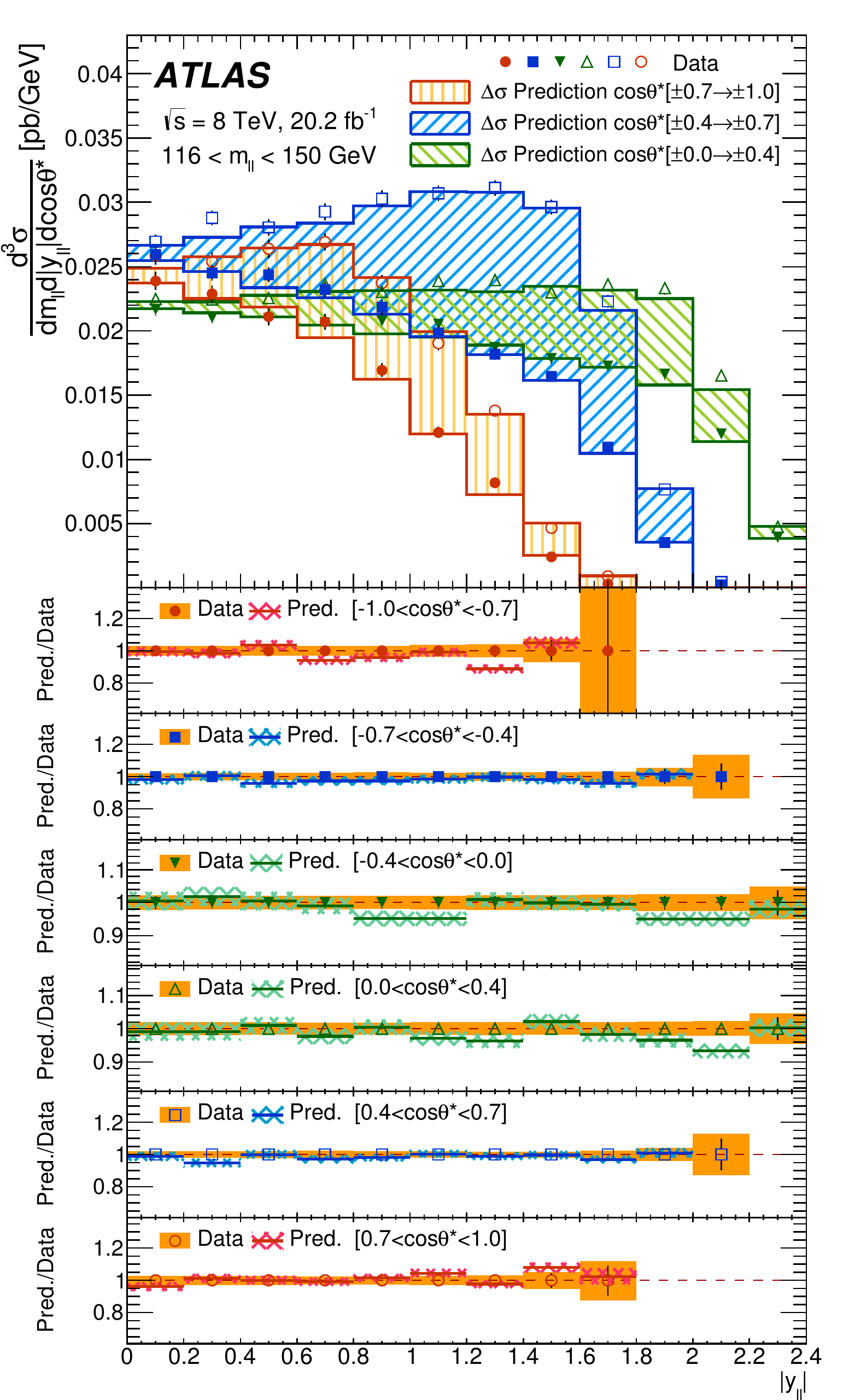}\\}
\caption{The combined 
Drell-Yan 
fiducial cross sections ${\rm d}^{3}\sigma$,
triply 
differential in dilepton invariant mass, dilepton rapidity and 
Collins-Soper angle, measured by ATLAS.  See text for details, Figure from Ref.~\cite{ATLAS:2017rue}. }
\label{fig6}
\end{figure}
More differential 
information, especially on PDFs, is gained from the production of W+jet final states~\cite{ATLAS:2014fjg}, and from ratios between 
W+jet and Z+jet production, with ATLAS and CMS measurements~\cite{CMS:2011jak,ATLAS:2014nxa}
at central rapidity being complemented, here again, by 
LHCb~\cite{LHCb:2016nhs} in the forward region. 

The Born-level kinematics of Drell-Yan lepton 
pair production is determined by three variables: 
the di-lepton invariant mass and rapidity and 
the scattering angle in the centre-of-momentum 
system (Collins-Soper angle $\theta^\star$). Triply-differential 
distributions in these variables can be measured 
to high precision, as illustrated 
by the ATLAS measurement~\cite{ATLAS:2017rue}
in Figure~\ref{fig6} for data in the Z boson mass region (left frame) and above (right frame). 
The kinematic region shown is labelled in each plot. The data are shown as solid $({\rm cos}\,\theta^\star<0)$ and open $({\rm cos}\,\theta^\star>0)$ marker. They are compared to prediction including NNLO QCD~\cite{fewz} and NLO EW~\cite{Alioli:2010xd} K-factors, shown as solid line. The difference, $\Delta\sigma$, between the predicted cross sections in the two measurement bins at equal $|{\rm cos}\,\theta^\star|$ symmetric around ${\rm cos}\,\theta^\star=0$ is represented by the hatched shading. In each plot, the lower panel shows the ratio of prediction to measurement. The inner error bars represent the statistical uncertainty of the data and the solid band	shows the total experimental uncertainty. The contribution to the uncertainty from the luminosity measurement is not included. The cross-hatched band represents the statistical and PDF uncertainties in the prediction.

These triply differential 
cross section data can be represented alternatively as angular coefficients~\cite{ATLAS:2016rnf} or 
be used to extract the forward-backward cross section asymmetry
in Drell-Yan production. 
Especially the forward-backward asymmetry 
is highly sensitive on the electroweak mixing angle,
and has been used for its precision determination~\cite{CMS:2018ktx,ATLAS:2018gqq,LHCb:2015jyu}. 

A hadron collider 
measurement of the W boson mass is essentially based on 
precision measurements of the lepton transverse momentum distribution, like the one exemplified in Figure~\ref{fig5} (bottom), 
in events with missing transverse momentum. For the W mass determination, a highly accurate prediction 
of the W boson transverse momentum distribution is 
crucial. It is obtained in a data-driven manner from the measured Z boson transverse momentum distribution, which is then combined with precision theory predictions~\cite{fewz}, now available even at NNLO+N3LL~\cite{vresum} for the W/Z ratio. 
The low luminosity data, collected at center-of-mass energies of 5 and 13 TeV during the Run 2 of the LHC, will allow to measure the W transverse momentum distribution and hence further constrain the available predictions. The W mass is then determined from the shape
of the measured lepton transverse momentum distribution and the transverse mass distribution. It reaches an accuracy of 
19~MeV at ATLAS~\cite{ATLAS:2017rzl} and 
31 MeV at LHCb~\cite{LHCb:2021mW}. 
The upgraded ATLAS and CMS detectors for the high luminosity phase will have enhanced tracking acceptance, which will allow for extended measurements towards the forward region. Combined with some dedicated low luminosity runs, this will allow to further enhance the precision of these measurements.

Production of electroweak gauge bosons in 
association with heavy flavour jets 
can help to constrain specific quark PDFs,
such as the strange quark distribution form 
W+charm production~\cite{LHCb:2016hnm}. Photon or 
Z boson production 
in association with charm or bottom quark jets~\cite{ATLAS:2017qlc} directly probes the respective PDFs and has sensitivity on the gluon fragmentation to bottom quark pairs.

\section{Extractions of parton distributions}
\label{sec:pdf}

Protons are complex bound states of quarks, held together by gluons. Their bound state dynamics is governed by low-energy QCD, corresponding to the limit of a large coupling constant where perturbation theory is not applicable.  The concept of factorization in QCD does however allow to separate these long-distance bound-state dynamics from the short-distance dynamics of hard scattering processes of quarks and gluons, by introducing parton distribution functions (PDFs) that describe the probability distributions for all flavours of quarks and anti-quarks as well as for gluons inside the proton 
as function of their momentum fraction $x$ and of the resolution scale $Q$ (factorization scale) at which the proton is probed.   
The theory predictions for proton-proton collider observables are then obtained by convoluting the hard scattering cross sections for parton-parton interactions with the PDFs describing the distributions of partons inside the colliding protons. This convolution is performed at the factorization scale $Q$ that is characteristic to the final state under consideration.

Perturbative QCD predicts the dependence of these PDFs as a function of $Q$, in the form of the DGLAP evolution equations~\cite{dglap}, which form a set of first-order integro-differential equations with perturbatively calculable kernels (splitting functions). These splitting functions describe the parton-to-parton transition probabilities as functions of the momentum fraction in the transition. They can be computed order-by-order in perturbative QCD, and they are currently known to three loops~\cite{mvv}, which corresponds to NNLO accuracy.  
To solve the DGLAP evolution equations, initial conditions for the PDFs must be specified in the form of the $x$-dependent distributions at some initial scale $Q_0$. These initial conditions contain the information on the non-perturbative bound state dynamics, which can not be calculated from first principles with current methodology. 

Instead, one uses data on a variety of hard scattering processes from different particle colliders to constrain the different PDF initial conditions through a global fit.  The basic methodology of these PDF fits is as follows. All PDFs are parametrized 
as functions of $x$ at the initial scale $Q_0$ 
in terms of a limited set of a priori unknown parameters. For fixed values of the parameters, the 
PDFs are evolved by solving the DGLAP evolution equations (truncated to the desired perturbative order), yielding PDFs for all values of $Q$. These are then used to compute the cross section predictions for all sets of observables that are included in the fit, expressing the parton-parton cross sections at the same perturbative order as used in the evolution. By comparing the resulting predictions with the available data for all observables, the quality of the description can be quantified.
The procedure is repeated in an iterative manner until the fit converges onto an optimum set of parameters (and determines the associated uncertainties). 

Global PDF fits are performed by several different groups: ABM~\cite{Alekhin:2017kpj}, CTEQ~\cite{Hou:2019efy}, MSHT~\cite{Bailey:2020ooq} and NNPDF~\cite{NNPDF:2021}, which differ in the selection of data sets included in the fit, in the form of the parametrizations at the initial scale, in the treatment of heavy quark mass effects, as well as in the
handling of experimental correlations and in the procedures used to quantify the uncertainty on the fitted PDFs. Several other PDF fits are performed on restricted data sets. 
While in a series of experimental publications~\cite{ATLAS:2017ble,ATLAS:2017kux,ATLAS:2013jmu} quantitative data-theory comparisons were performed using both the experimental and theoretical uncertainties (with their correlations between different phase-space regions), PDF fits traditionally used experimental uncertainties only. As a fruit of the intense interactions between experimentalists and theorists, in particular in the context of the PDF4LHC forum, theoretical uncertainties are being considered for the PDF fits~\cite{Ball:2021icz}. In recent global PDF fits, the impact of the uncertainties on the correlations, first discussed in Refs.~\cite{ATLAS:2017ble,ATLAS:2017kux}, is starting to be quantified~\cite{Hou:2019efy,Bailey:2020ooq,NNPDF:2021}. It was observed that while correlations do have important consequences for the fit quality, the impact is rather moderate for the resulting PDFs themselves. 
\begin{figure}[t]
\parbox{0.5\columnwidth}{\includegraphics[width=0.48\columnwidth]{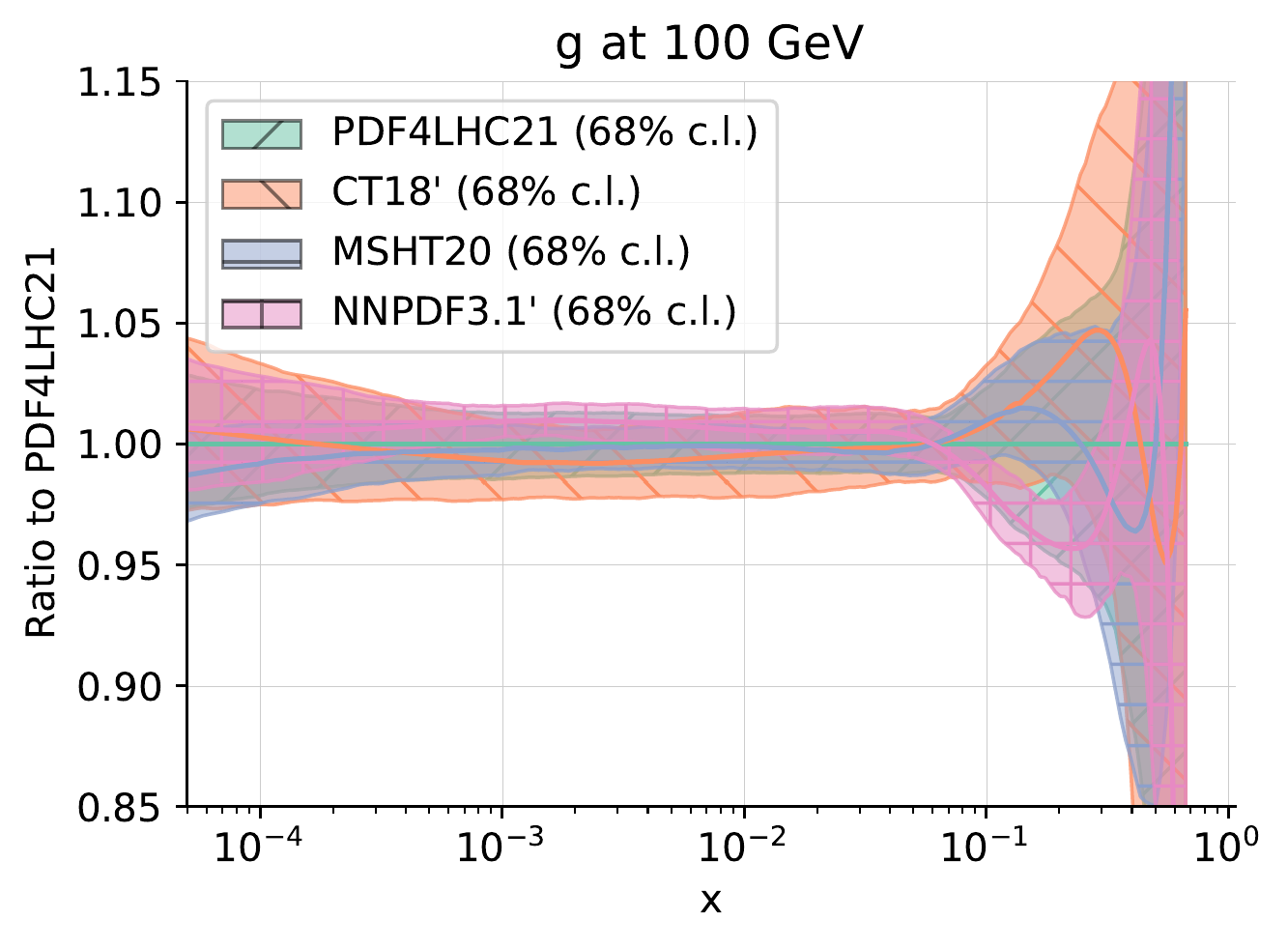}}
\parbox{0.5\columnwidth}{\includegraphics[width=0.48\columnwidth]{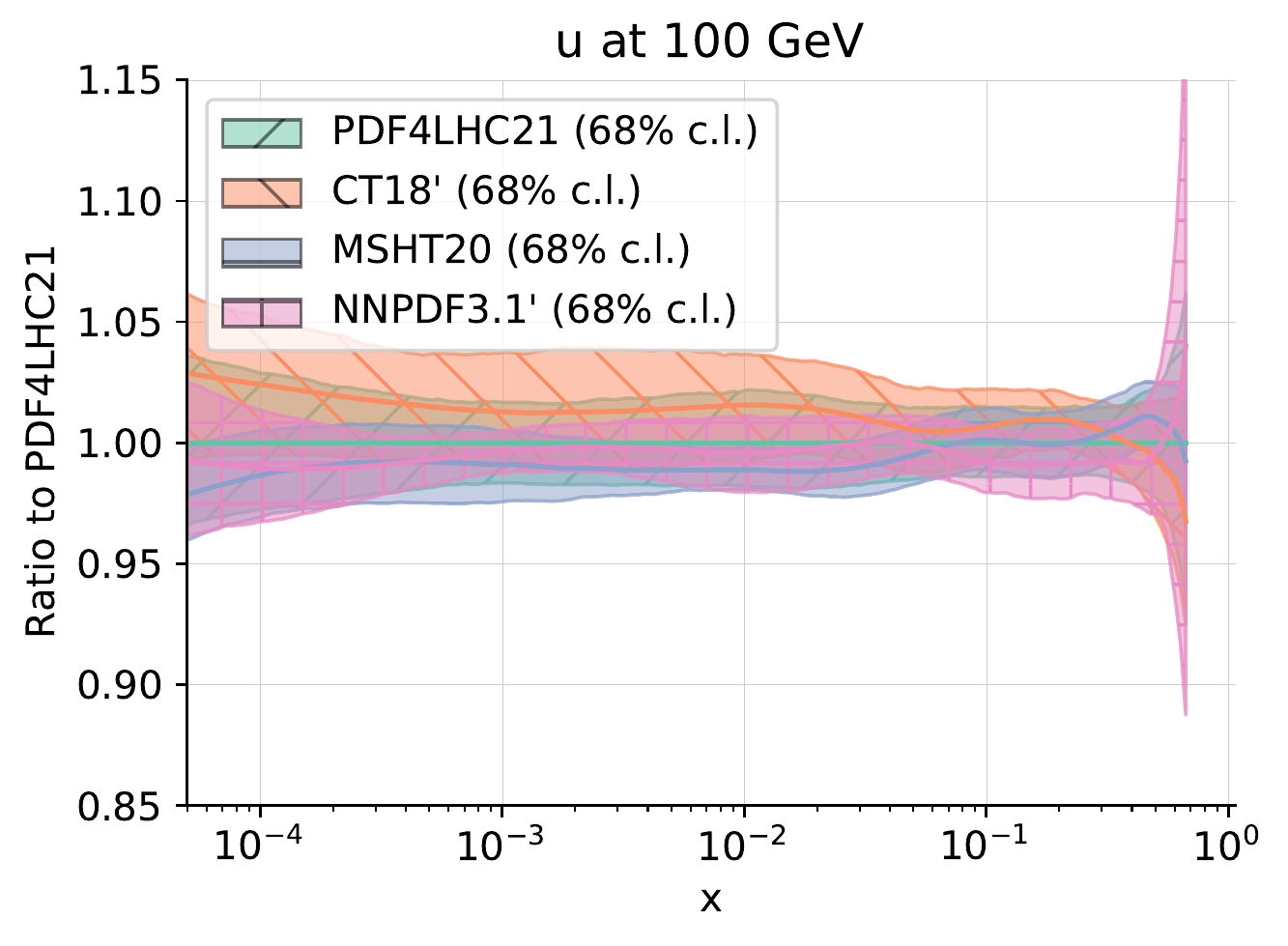}} \\
\parbox{0.5\columnwidth}{\includegraphics[width=0.48\columnwidth]{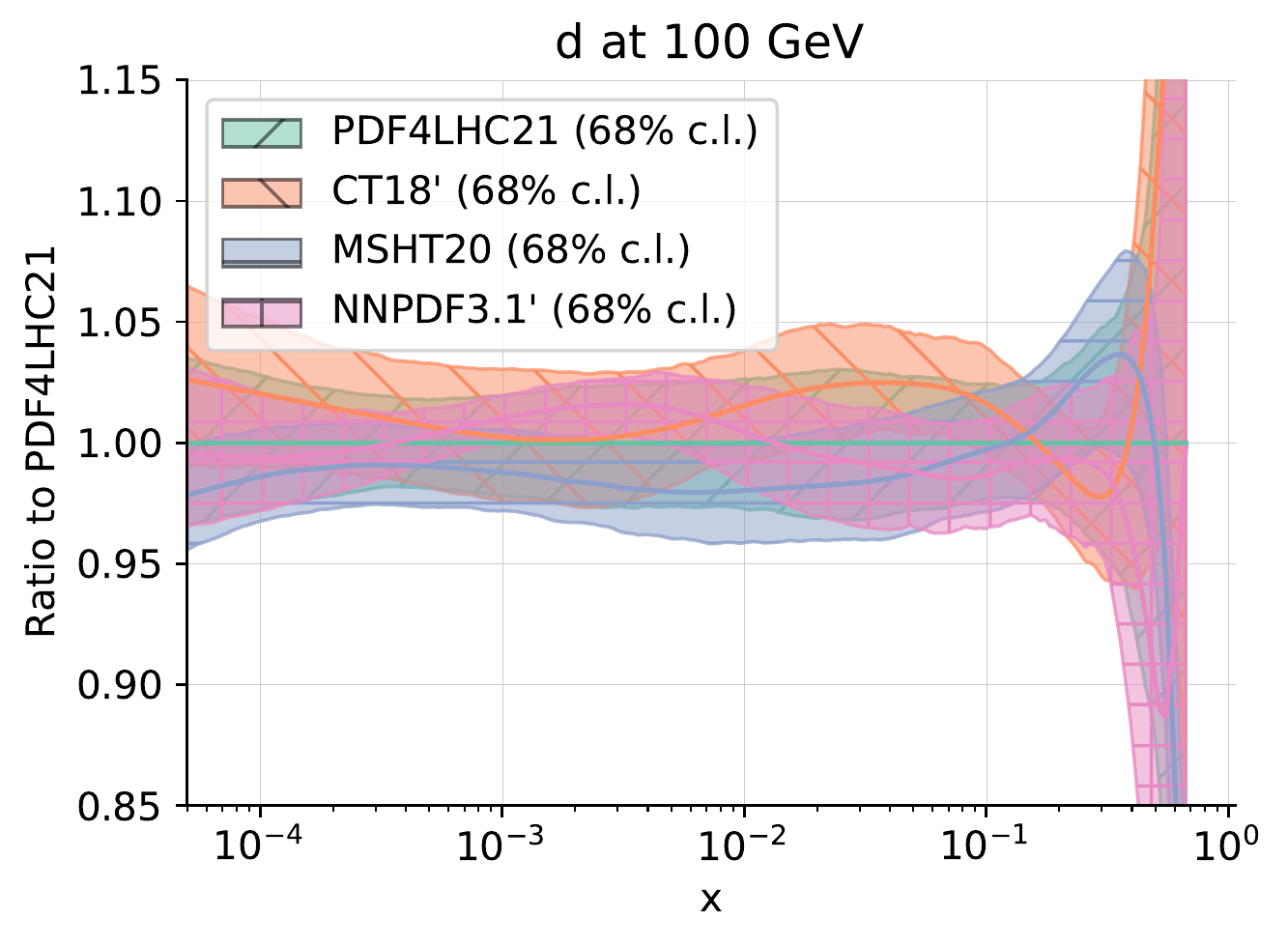}}
\parbox{0.5\columnwidth}{\includegraphics[width=0.48\columnwidth]{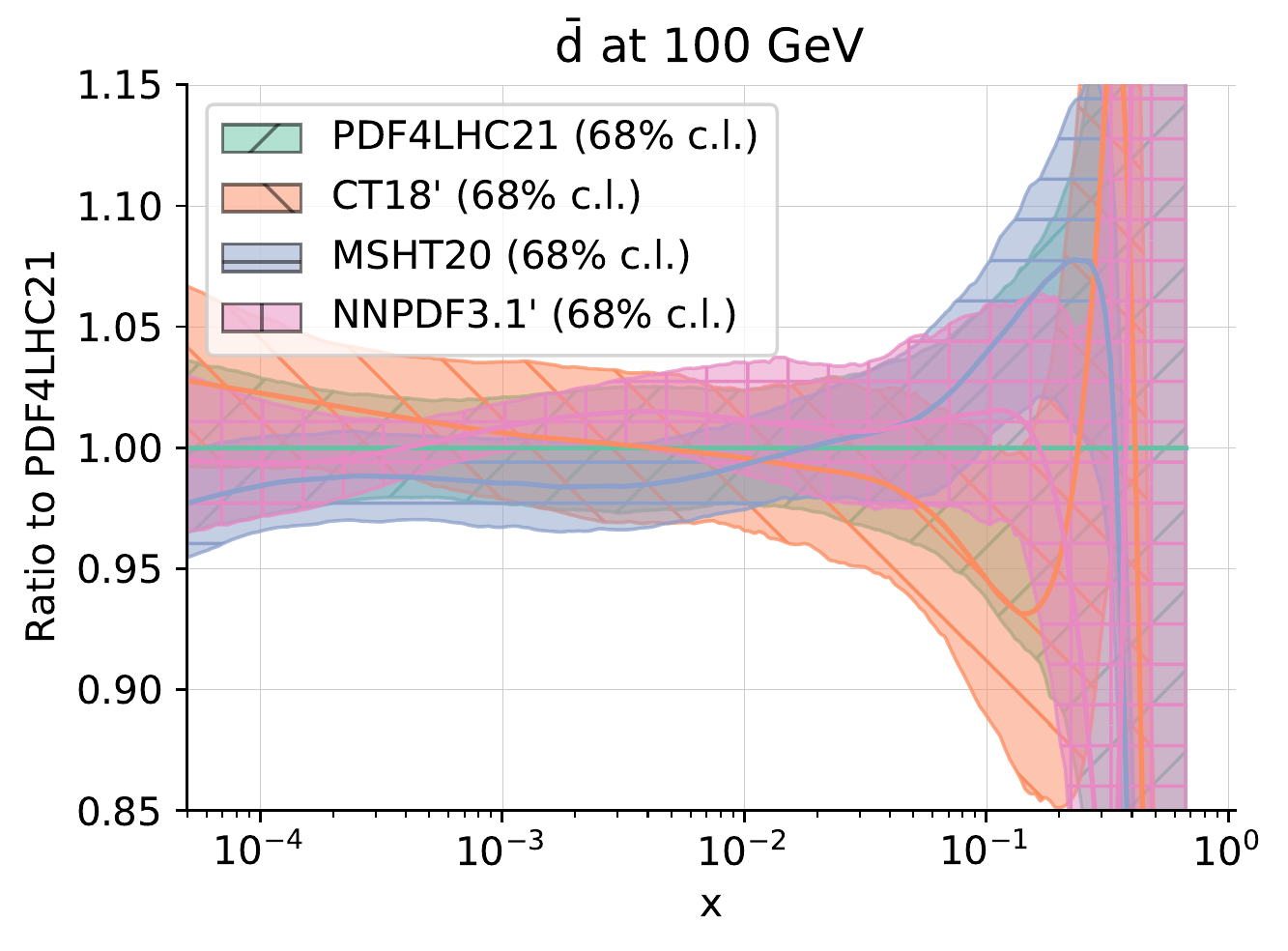}}
\vspace{0.5cm}
\caption{Comparison of the PDF4LHC21 combination (composed of $N_{\rm rep}=900$ replicas) with the three constituent sets at $Q=100$ GeV, normalised to the central value of the former and with their respective 68\%CL uncertainty bands. In the case of the Hessian sets (${\rm CT18}^\prime$ and MSHT20) their Monte Carlo representation composed of $N_{\rm rep}=300$ replicas is displayed. The ${\rm NNPDF3.1}^\prime$ band is also constituted by $N_{\rm rep}=300$ (native) replicas. The ${\rm g}$, ${\rm u}$, ${\rm d}$ and $\bar{\rm d}$ components are displayed. Figure taken from Ref.~\cite{Ball:2022hsh}.
}
\label{fig8}
\end{figure}

A dominant source of information on the quark and antiquark PDFs are structure functions from deep inelastic scattering (DIS) experiments of leptons or neutrinos on protons or light nuclei, which probe different linear combinations of the quark distributions. With the exception of structure functions from the DESY HERA electron-proton collider, these data are mostly taken at fixed target experiments with relatively low values of $Q$, potentially inducing sensitivity on target mass corrections, and cover only momentum fractions above $x\approx 0.05$. Precision measurements of vector boson observables at the LHC are now starting to provide complementary constraints, which reach to lower values of $x$ and probe novel combinations of the quark distributions. Combined with the HERA data, it can be expected that future LHC vector boson data will become sufficient to supersede the fixed-target DIS data.  

The gluon distribution can be probed through inclusive jet and di-jet production, as well as from the transverse momentum distributions of massive gauge bosons and photons, or from top quark pair production. 
All these collider processes have a sensitivity on the gluon distribution already at leading order, which is 
in contrast to the DIS structure functions. Consequently, the main constraints on the gluon distribution derive from Tevatron and LHC data. 
Optimal sensitivity to quark and gluon distributions is however reached only through a combined fit of multiple observables, which are closely interlinked through the DGLAP evolution equations and mutually constraining each other even at the initial scale through the requirement of proton momentum conservation. 

The PDF4LHC working group~\cite{Butterworth:2015oua} regularly benchmarks the 
results of the different global fits in order to identify and explain mutual deviations, to issue recommendations for the uncertainties that should be associated with the variations among different PDF sets. It also produces prescriptions for the merging of the results of different global fits, with the PDF4LHC21 update released recently~\cite{Ball:2022hsh}.  

In Figure~\ref{fig8}, the results of the CT18~\protect\cite{Hou:2019efy}, NNPDF3.1~\protect\cite{NNPDF:2017mvq} and MSHT20~\protect\cite{Bailey:2020ooq} PDF fits, together with their uncertainties, are compared with the result of their statistical merging performed within the PDF4LHC21 exercise~\cite{Ball:2022hsh}. 
It is observed that the three sets are mutually consistent within their quoted uncertainties. 
In the phenomenologically interesting $x$ range, the various PDF components are known to few~\% or better. 
For high-$x$ the PDF uncertainties increase rapidly. 
It is especially in this region that future HL-LHC precision measurements will provide unique new PDF information. 
A similar deterioration of the uncertainties is also seen at very small $x$, which could be probed by dedicated precision measurements in the very forward region.

The strong coupling constant $\alpha _\mathrm {s}$ can be determined as part of the PDF fit, usually resulting in values consistent with the world average albeit with larger uncertainty. However, the resulting parton distributions can then be used in a consistent manner only together with this fixed value of $\alpha _\mathrm {s}$, thus making them unsuitable for $\alpha _\mathrm {s}$ determinations from collider observables. As an alternative, most fits provide families of PDFs with different values of $\alpha _\mathrm {s}$, or provide prescriptions to correct their default PDF set for the effects of $\alpha _\mathrm {s}$ variations. 
The consistent inclusion of these effects 
is particularly important for $\alpha _\mathrm {s}$
determinations from fully inclusive observables, such as the Drell-Yan type processes~\cite{CMS:2019oeb}.  It is less of an issue if $\alpha _\mathrm {s}$ 
is determined from jet cross-section ratios~\cite{CMS:2013vbb} or 
normalized event shape distributions~\cite{ATLAS:2017qir}, which were discussed above.

\section{QCD implications for LHC measurements and searches}
\label{sec:searches}

The search for signatures of physics 
beyond the Standard Model (BSM) is a main 
objective of the LHC physics program. 
Broadly speaking, one distinguishes direct searches for the production of new particles, observed through distinctive final state signatures, and indirect searches, which aim to uncover small 
deviations from Standard Model expectations in precision observables. For both types of searches, precision QCD plays a crucial role in converting experimental measurements into constraints on BSM models.  
\begin{figure}[t]
\parbox{0.45\columnwidth}{\includegraphics[width=0.45\columnwidth]{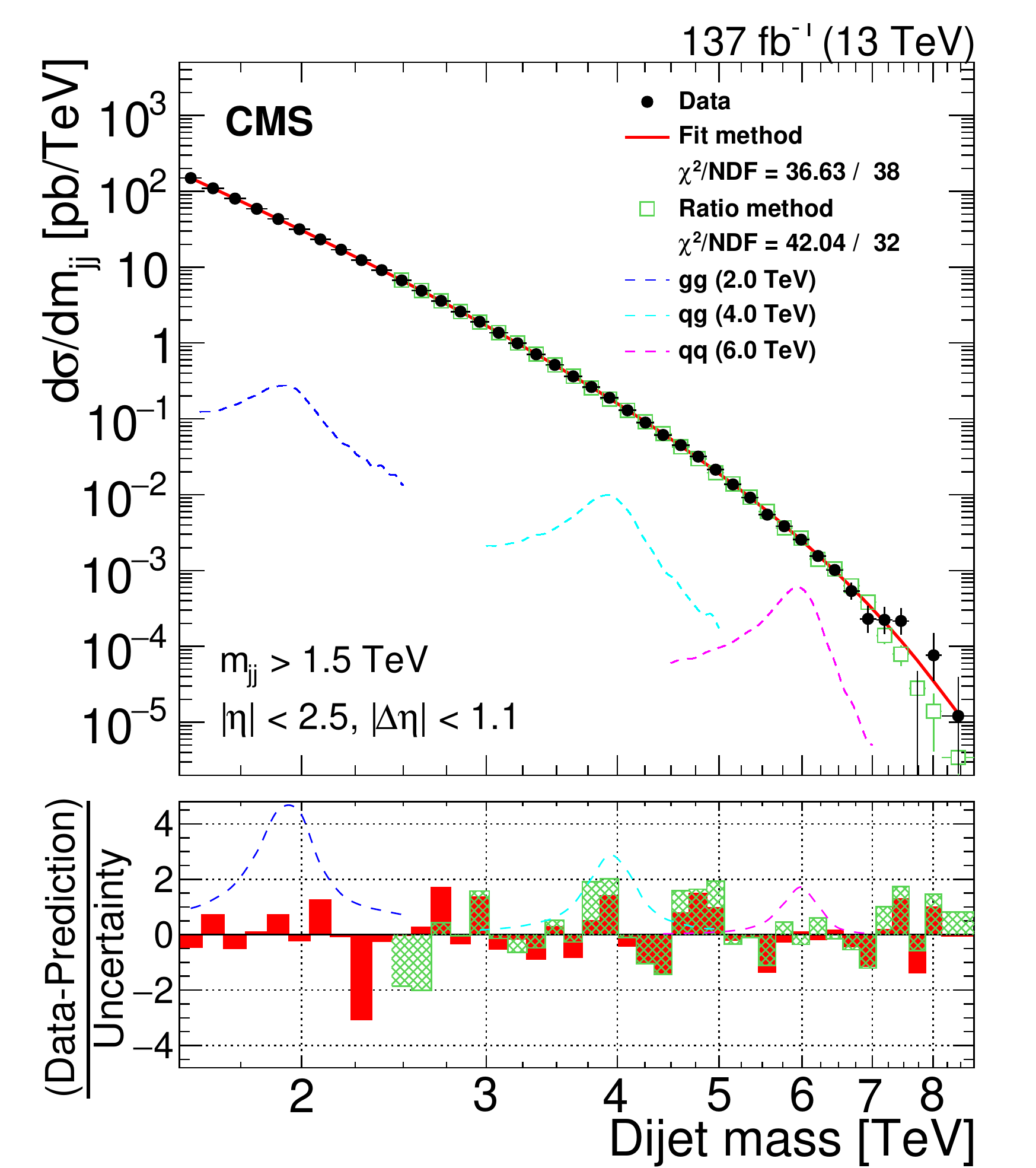}\\}
\parbox{0.55\columnwidth}{\includegraphics[width=0.55\columnwidth]{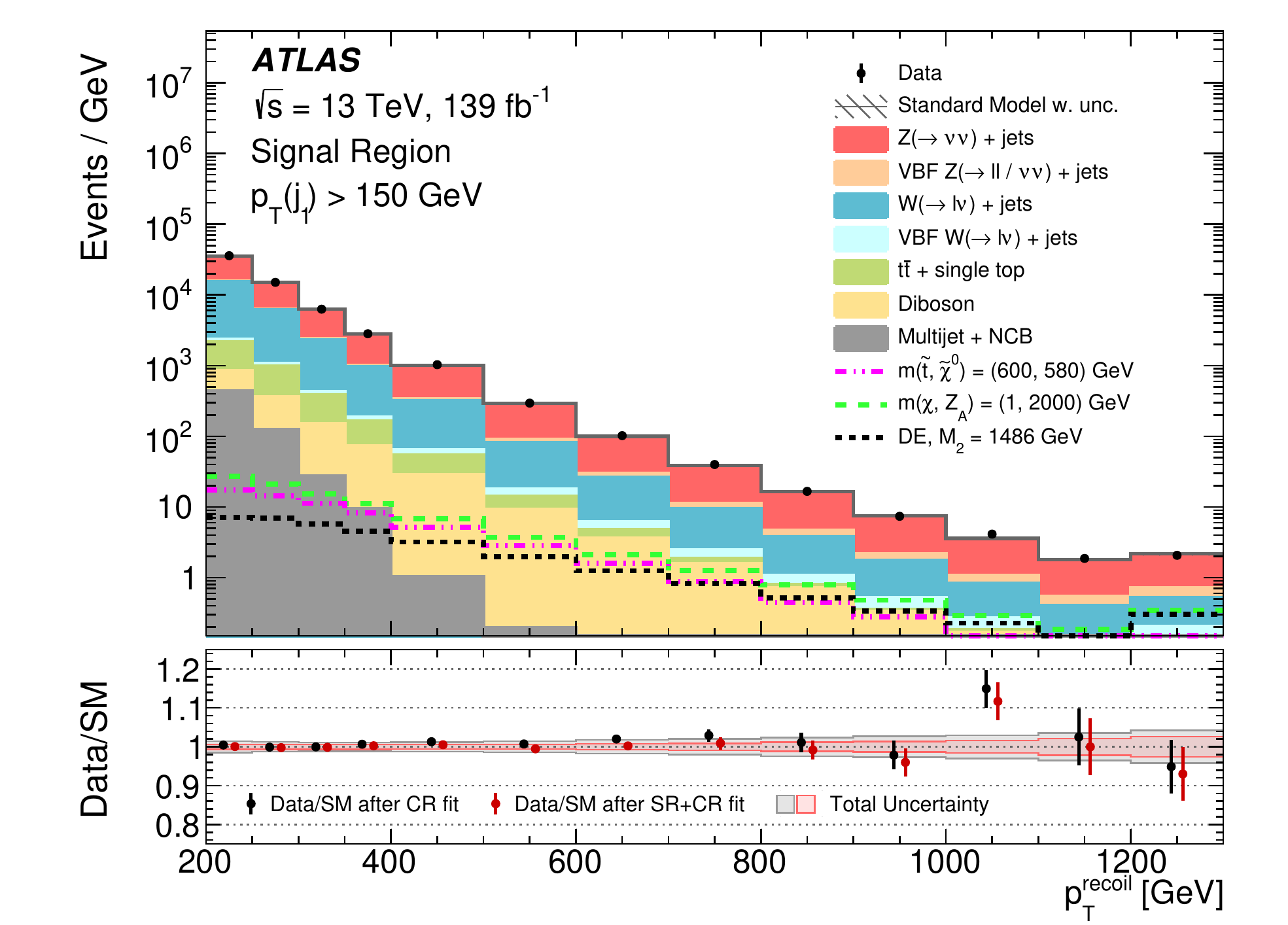}}
\caption{ Left: Dijet mass spectrum measured by CMS 
used to establish cross section limits on 
the production of di-jet resonances, see text for details. Figure taken from~\cite{CMS:2019gwf}.
Right: Distribution of $p_{\rm T}^{\rm recoil}$
in jet-plus-missing energy BSM search, measured by ATLAS for the $p_{\rm T}^{\rm recoil} >200 $ GeV selection and compared with the SM predictions. See text for details. Figure taken from Ref.~\cite{ATLAS:2021kxv}.
}
\label{fig7}
\end{figure}

Direct searches aim to discover BSM particles through the detection of their decay signatures. In an ideal setting, such a particle can be observed as a peak in a mass distribution over a continuum background, 
as for example in searches for di-jet 
resonances~\cite{ATLAS:2017eqx,CMS:2019gwf} in 
invariant mass spectra or angular distributions. 
Figure~\ref{fig7} (left)  shows the 
di-jet mass spectrum measured by CMS~\cite{CMS:2019gwf} in the signal region (points) compared to a fitted parameterization of the background (solid line) and the one obtained from the control region (green squares). Examples of predicted signals from narrow gluon-gluon, quark-gluon, and quark-quark resonances are shown (dashed coloured lines) with cross sections equal to the observed upper limits at $95\%$ confidence level. 
An alternative approach consists in using unfolded cross-section measurements, to be compared with SM and BSM particle-level predictions~\cite{ATLAS:2013jmu,CMS:2018ucw}, with the practical advantage that the convolution with the full detector simulation is not necessary for the various models that are being tested.

In most BSM searches, decay signatures are more complex, can involve unseen particles (missing energy) and may suffer from a limited resolution in the mass reconstruction. In these cases, an accurate description of the cross sections and associated uncertainties for all SM background processes that can produce the final state under consideration becomes crucial. Especially in the high-mass range, PDF uncertainties can become a limiting factor, see also Figure~\ref{fig8}.

These background predictions are often obtained through data-driven methods with some input from theory. Theory predictions from multi-purpose event generators, corrected for higher-order QCD effects, are validated and normalised on control regions where no signal is expected, and then extrapolated to the signal regions where they are compared to observations.
Searches involving jets-plus-missing-energy signatures are classical examples~\cite{ATLAS:2021kxv,CMS:2021far}. 
Figure~\ref{fig7} (right) displays the distribution of the missing transverse momentum 
$p_{\rm T}^{\rm recoil}$ measured by ATLAS~\cite{ATLAS:2021kxv} for 
jet-plus-missing-energy events with $p_{\rm T}^{\rm recoil} >200 $ GeV compared with the SM predictions in the signal region.
The latter are normalized with normalization factors as determined by the global fit that considers exclusive $p_{\rm T}^{\rm recoil}$ control regions (``CR fit"). For illustration purposes, the distributions of examples of dark energy (DE), SUSY, and WIMP scenarios are included. The ratios of data to SM predictions after the CR fit are shown in the lower panel (black dots), and compared with the same quantities when SM predictions are normalized to the results of the global background-only fit when the signal region is also included (``SR+CR fit", red dots). The error bands in the ratio shown in the lower panel include both the statistical and systematic uncertainties in the background predictions. Events with values beyond the range of the histogram are included in the last bin. 

Indirect searches for new physics are made with 
practically all precision observables at the LHC. Detailed comparisons of measurement results with theory predictions represent precision tests of the Standard Model at the quantum level.  The quantitative power of these precision tests is determined by the uncertainties on the experimental data and on the theory predictions. Especially for low-multiplicity collider processes, these uncertainties will be reaching per-cent level or below with the upcoming HL-LHC data set and with advances in N3LO QCD calculations.

Combinations of measurements can then be used for precision  determinations of Standard Model parameters, as for the combined measurements~\cite{ATLAS:2019nkf,CMS:2018uag} of Higgs boson production and 
decay modes, which turned into determinations of the Higgs boson couplings to the Standard Model particles. Subsequent combinations across experiments~\cite{ATLAS:2016neq} are then forming part of the LHC physics legacy results. 
These Higgs boson studies highlight the close interplay between experiment and theory on precision observables. With the help of simulations and theory predictions, the experimental data for fiducial cross sections are converted into pseudo-observables which can then be used to combine measurements from different observables (production processes or decay channels) to enhance the measurement sensitivity. Parameter extractions are then performed through data-theory comparisons at the level of these pseudo-observables. The large size of perturbative QCD corrections to many Higgs observables (e.g.\ a factor two increase of the gluon fusion cross section prediction from NLO QCD corrections) mandates the inclusion of higher order effects at all stages of this procedure. QCD-related uncertainties from missing higher-orders, parton distributions or QCD parameters contribute to the error budget of the parameter extractions, and progress on QCD will in turn help to improve the precision of LHC results in the Higgs sector and in all other aspects of Standard Model physics.

A very important class of indirect searches are 
precision studies of rare decays of heavy quarks that
have yielded several tantalizing results in 
the recent past, see e.g.~\cite{LHCb:2021trn}, which  are however beyond the scope of this report.  

To quantify the sensitivity on BSM effects that can be 
obtained from specific observables (precision measurements or probes of rare processes)
or from their combination, a model-agnostic approach is provided by the framework of Standard Model effective field theories, SMEFT,~\cite{Brivio:2017vri}.
In SMEFT, a basis of independent operators, composed from the Standard Model particle content, is considered at a fixed mass dimension. While this operator basis is very large already at dimension-6, one finds for individual observables that they receive contributions typically only from a small subset of basis elements.  
The coefficients of the operators can then be constrained from experimental data, either from specific observables, e.g.~\cite{ATLAS:2021dqb} or in a global fit~\cite{Ethier:2021bye}. The resulting operator coefficients can then be turned into estimates of the mass scales where the onset of BSM physics effects could be expected, or be confronted with specific BSM model predictions.

\section{Summary and Outlook}

In this review, we summarized the current status of precision QCD measurements on LHC data and discussed their implications for 
Standard Model precision tests and parameter determinations. We highlighted the multi-faceted nature of the precision QCD observables that can be defined on all-hadronic final states or in vector boson production processes. The interpretation of these data requires equally precise theory predictions, taking into account the detailed kinematical settings of the measurements. The ongoing NNLO-revolution, as well as advances in resummation and event generation, 
are starting to provide these precision theory predictions for an increasing number of observables. 

The next years of LHC running will substantially enlarge the LHC 
data set, 
thereby improving on statistical uncertainties and enabling access to rare processes, and provide opportunities for dedicated runs (e.g.\ low pileup) to mitigate sources of systematic uncertainty.

Achieving stronger experimental constraints on theoretical predictions through precision measurements will have to go hand-in-hand with improvements in the understanding of detector performance.
These become challenging when entering the area of sub-percent level uncertainties, as many subtle effects need to be understood and kept under control.
In particular, one has to be able to disentangle detector and physics effects for the in-situ calibration procedures, as well as when performing the physics measurements.
Strong interplay between the experimental and theoretical communities will be required in order to achieve the required detailed  understanding of physics modelling effects, which are very relevant in these studies.

In addition to the benefits from the upgraded detectors discussed above, the large data samples that will be collected in the high luminosity phase of the LHC will not only boost the studies of rare processes, but will also enable binned multi-differential measurements.
Furthermore, recent developments in machine learning techniques enable unbinned many-dimensional measurements, which facilitate comparisons between experiments and with theoretical predictions, allowing also to define other observables after the measurements are performed~\cite{Arratia:2021otl}.
For example, such approaches may allow to simultaneously measure numerous event- and jet-level properties, opening a broad new area for probing subtle properties of QCD.
Doing so, one of the main challenges will be to achieve a good understanding of the full set of systematic uncertainties (experimental, theoretical and from the unfolding procedure itself), together with their correlations across the high dimensional phase-space of the measurement.

A comparable sub-per-cent precision on the theory predictions 
represents a major challenge in terms of obtaining higher-order 
predictions in fixed-order perturbation theory and resummation, which will require the development of novel algebraical and numerical approaches to handle the increasing complexity of the underlying 
field-theory expressions in an efficient and reliable way. Important first steps towards the automation of NNLO calculations, to 
the development of new methods for N3LO calculations and their application in specific processes, and to resummation that is commensurate with this 
accuracy level are currently being made. At the 
quantitative precision that is aimed for in QCD in the near future,  
non-perturbative effects in hadronization and proton structure will 
start to play an increasingly important role, thereby posing entirely novel types of challenges, and asking for new types 
of observables to probe them. The quantification of uncertainties on the theory predictions, currently approximated through scale variations, will also have to be revisited, aiming for 
a reliable theory error estimation with its proper statistical interpretation.

\section*{DISCLOSURE STATEMENT}
The authors are not aware of any affiliations, memberships, funding, or financial holdings that might be perceived as affecting the objectivity of this review. 

\section*{ACKNOWLEDGMENTS}
The authors would like to thank
Paolo Azzurri, Guillelmo Gomez Ceballos, Cvetan Cheshkov, Andrea Dainese, Monica Dunford, Massimiliano Grazzini, Alexander Kalweit, Evelin Meoni, Lorenzo Sestini, Carlos Vazquez for useful discussions, suggestions and comments on the manuscript. 

TG gratefully acknowledges funding from the European Research Council (ERC) under the European Union's Horizon 2020 research and innovation programme grant agreement 101019620 (ERC Advanced Grant TOPUP) and from the Swiss National Science Foundation (SNF) under contract 200020-204200. 

BM gratefully acknowledges the continuous support from LPNHE, CNRS/IN2P3, Sorbonne Universit\'e and Universit\'e de Paris.

%


\noindent


\end{document}